\documentclass[twocolumn]{aastex61}
\usepackage{natbib}
\newcommand{\Mstar}{\ifmmode M_{\ast} \else $M_{\ast}$\fi}
\newcommand{\kms}{\ifmmode \mathrm{km~s^{-1}} \else km~s$^{-1}$\fi}
\newcommand{\smpy}{\ifmmode M_{\sun}~\mathrm{yr}^{-1} \else $M_{\sun}$~yr$^{-1}$\fi}
\newcommand{\msol}{\ifmmode M_{\sun} \else $M_{\sun}$\fi}
\newcommand{\smpc}{\ifmmode M_{\sun}~\mbox{pc}^{-2} \else $M_{\sun}$~pc$^{-2}$\fi}
\newcommand{\smkpc}{\ifmmode M_{\sun}~\mbox{yr}^{-1}~\mbox{kpc}^{-2} \else $M_{\sun}$~yr$^{-1}$~kpc$^{-2}$\fi}
\newcommand{\htwo}{\ifmmode \text{H}_{2} \else H$_{2}$\fi}
\newcommand{\ha}{\ifmmode \mbox{H}\alpha \else H$\alpha$\fi}
\newcommand{\sightwo}{\ifmmode \Sigma_{\textnormal{H}_{2}} \else $\Sigma_{\textnormal{H}_{2}}$\fi}
\newcommand{\sigmol}{\ifmmode \Sigma_{\textnormal{mol}} \else $\Sigma_{\textnormal{mol}}$\fi}
\newcommand{\sigsfr}{\ifmmode \Sigma_{\textnormal{SFR}} \else $\Sigma_{\textnormal{SFR}}$\fi}
\newcommand{\siggas}{\ifmmode \Sigma_{\textnormal{gas}} \else $\Sigma_{\textnormal{gas}}$\fi}
\newcommand{\sighi}{\ifmmode \Sigma_\text{H\textsc{i}} \else $\Sigma_\text{H\textsc{i}}$\fi}
\newcommand{\sigdust}{\ifmmode \Sigma_\text{dust} \else $\Sigma_\text{dust}$\fi}
\newcommand{\sigstar}{\ifmmode \Sigma_{*} \else $\Sigma_{*}$\fi}
\newcommand{\sigdotstar}{\ifmmode \dot{\Sigma}_{*} \else $\dot{\Sigma}_{*}$\fi}
\newcommand{\cplus}{\ifmmode \mbox{C}^{+} \else C$^{+}$\fi}
\newcommand{\CII}{\ifmmode \text{[C\textsc{ii}]} \else {\sc [C\,ii]}\fi}
\newcommand{\thirteenCII}{\ifmmode \text{[}^{13}\text{C\textsc{ii}]} \else {\sc [$^{13}$C\,ii]}\fi}
\newcommand{\twelveCII}{\ifmmode \text{[}^{12}\text{C\textsc{ii}]} \else {\sc [$^{12}$C\,ii]}\fi}
\newcommand{\CI}{\ifmmode \text{[C\textsc{i}]} \else {\sc [C\,i]}\fi}
\newcommand{\OI}{\ifmmode \text{[O\textsc{i}]} \else {\sc [O\,i]}\fi}
\newcommand{\NII}{\ifmmode \text{[N\textsc{ii}]} \else {\sc [N\,ii]}\fi}
\newcommand{\OIII}{\ifmmode \text{[O\textsc{iii}]} \else {\sc [O\,iii]}\fi}

\newcommand{\hi}{\ifmmode \text{H\textsc{i}} \else {\sc H\,i}\fi}
\newcommand{\hii}{\ifmmode \text{H\textsc{ii}} \else H\,{\sc ii}\fi}
\newcommand{\Nhi}{\ifmmode N_\text{H\textsc{i}} \else $N_{\text{H\textsc{i}}}$\fi}
\newcommand{\Nmol}{\ifmmode N_\text{mol} \else $N_\text{mol}$\fi}
\newcommand{\Nhtwo}{\ifmmode N_{\textnormal{H}_{2}} \else $N_{\textnormal{H}_{2}}$\fi}
\newcommand{\Mhi}{\ifmmode M_\text{H\textsc{i}} \else $M_\text{H\textsc{i}}$\fi}
\newcommand{\Mhtwo}{\ifmmode M_{\textnormal{H}_{2}} \else $M_{\textnormal{H}_{2}}$\fi}
\newcommand{\Mmol}{\ifmmode M_{\textnormal{mol}} \else $M_{\textnormal{mol}}$\fi}
\newcommand{\dGDR}{\ifmmode \delta_{\textnormal{GDR}} \else $\delta_{\textnormal{GDR}}$\fi}
\newcommand{\CO}{\ifmmode ^{12}\rm{CO} \else $^{12}$CO\fi}
\newcommand{\thirteenCO}{\ifmmode ^{13}\rm{CO} \else $^{13}$CO\fi}
\newcommand{\CeighteenO}{\ifmmode \rm{C}^{18}\rm{O} \else C$^{18}$O\fi}
\newcommand{\taudep}{\ifmmode \tau_{\textnormal{dep}} \else $\tau_{\textnormal{dep}}$\fi}
\newcommand{\taumoldep}{\ifmmode \tau^{\textnormal{mol}}_{\textnormal{dep}} \else $\tau^{\textnormal{mol}}_{\textnormal{dep}}$\fi}
\newcommand{\Av}{\ifmmode A_{V} \else $A_{V}$\fi}
\newcommand{\COtoCII}{\ifmmode \textnormal{CO}/\CII \else $\textnormal{CO}/\CII$}

\revised{\today}

\shorttitle{ATCA \hi\ Absorption Survey of the SMC}
\shortauthors{Jameson et al.}

\begin{document}

\title{An ATCA Survey of \hi\ Absorption in the Magellanic Clouds I: \\ 
\hi\ Gas Temperature Measurements in the Small Magellanic Cloud}

\correspondingauthor{Katherine E. Jameson}
\email{katie.jameson@anu.edu.au}

\author[0000-0001-7105-0994]{Katherine E. Jameson}
\affil{Research School of Astronomy \& Astrophysics, Australian National University, Canberra, ACT 2611, Australia}

\author[0000-0003-2730-957X]{N. M. McClure-Griffiths}
\affiliation{Research School of Astronomy \& Astrophysics, Australian National University, Canberra, ACT 2611, Australia}

\author{Boyang Liu}
\affiliation{International Centre for Radio Astronomy Research (ICRAR), University of Western Australia, Crawley, WA 6009, Australia}
\affiliation{University of Chinese Academy of Sciences, Beijing 100049, People’s Republic of China}

\author[0000-0002-6300-7459]{John M. Dickey}
\affiliation{School of Natural Sciences, Private Bag 37, University of Tasmania, Hobart, TAS 7001, Australia}

\author[0000-0002-8057-0294]{Lister Staveley-Smith}
\affiliation{International Centre for Radio Astronomy Research (ICRAR), University of Western Australia, Crawley, WA 6009, Australia}
\affiliation{ARC Centre of Excellence for All Sky Astrophysics in 3 Dimensions (ASTRO 3D), Australia}

\author{Sne\u{z}ana Stanimirovi\'c}
\affiliation{Department of Astronomy, University of Wisconsin, Madison, WI 53706, USA}

\author[0000-0002-4899-4169]{James Dempsey}
\affiliation{Research School of Astronomy \& Astrophysics, Australian National University, Canberra, ACT 2611, Australia}
\affiliation{CSIRO Information Management and Technology, PO Box 225, Dickson, ACT 2602, Australia}

\author{J. R. Dawson}
\affiliation{Department of Physics and Astronomy and MQ Research Centre in Astronomy, Astrophysics and Astrophotonics, Macquarie University, NSW 2109, Australia}

\author[0000-0002-9214-8613]{Helga D{\'e}nes}
\affiliation{ASTRON, Netherlands Institute for Radio Astronomy, Oude Hoogeveensedijk 4, 7991 PD, Dwingeloo, The Netherlands}

\author{Alberto D. Bolatto}
\affiliation{Department of Astronomy, University of Maryland, College Park, MD 20742}

\author[0000-0003-3010-7661]{Di Li}
\affiliation{CAS Key Laboratory of FAST, NAOC, Chinese Academy of Sciences, People’s Republic of China}
\affiliation{University of Chinese Academy of Sciences, Beijing 100049, People’s Republic of China}

\author[0000-0002-7759-0585]{Tony Wong}
\affiliation{Department of Astronomy, University of Illinois, Urbana, IL 61801}



\begin{abstract}

We present the first results from the Small Magellanic Cloud portion of a new Australia Telescope Compact Array (ATCA) \hi\ absorption survey of both of the Magellanic Clouds, comprising over 800 hours of observations. Our new \hi\ absorption line data allow us to measure the temperature and fraction of cold neutral gas in a low metallicity environment. We observed 22 separate fields, targeting a total of 55 continuum sources against 37 of which we detected \hi\ absorption; from this we measure a column density weighted mean average spin temperature of $<T_{s}>=150$ K. Splitting the spectra into individual absorption line features, we estimate the temperatures of different gas components and find an average cold gas temperature of $\sim{30}$ K for this sample, lower than the average of $\sim{40}$ K in the Milky Way. The \hi\ appears to be evenly distributed throughout the SMC and we detect absorption in $67\%$ of the lines of sight in our sample, including some outside the main body of the galaxy (\Nhi$>2\times{10^{21}}$ cm$^{-2}$). The optical depth and temperature of the cold neutral atomic gas shows no strong trend with location spatially or in velocity. Despite the low metallicity environment, we find an average cold gas fraction of $\sim{20\%}$, not dissimilar from that of the Milky Way.

\end{abstract}




\section{Introduction} \label{sec:intro}

Atomic hydrogen gas (\hi) is typically the principal gas component within a galaxy. The ability of a galaxy to form cold \hi\ is a key rate-limiting step in its ability to form stars as it will regulate the ability to form molecular gas. The \hi\ is expected to be split between a warm and a cool phase in pressure equilibrium  (e.g., \citealt{mck77,wol95,wol03}) with the fractions and temperatures of the warm neutral medium (WNM) and cold neutral medium (CNM) depending on the exact balance between heating and cooling processes in the gas. For typical pressure ranges, the CNM is expected to have kinetic temperatures in the range of T$_{k}\sim{40-200}$ K and volume densities $n \sim{10-100}$ cm$^{-3}$ and for the WNM T$_{k}\sim{4000-8000}$ K and $n\sim10^{-2}-1$ cm$^{-3}$ for conditions found in the Milky Way \citep{wol03}. Observation of the 21-cm \hi\ absorption against background radio continuum sources is the primary way to directly trace the cold and/or optically thick \hi\ in a galaxy (e.g., \citealt{hei03a, mur15}). Comparing absorption to nearby \hi\ emission, one can measure the excitation temperature, or spin temperature, of \hi\ gas. 

Understanding the properties of the neutral gas at low metallicity is crucial for understanding how galaxies regulate the formation of molecular gas and stars. The Small Magellanic Cloud (SMC), with a metallicity of $\sim{1/5}$ solar metallicity \citep{duf84,kur99,pag03}, is one of the nearest gas-rich star-forming galaxies. It provides an ideal location to study the distribution and temperature of the \hi\ gas in a low metallicity environment. The close proximity of the SMC means it subtends a large angular area on the sky, making it possible to find a sufficient number of background radio sources to study the \hi\ absorption in detail throughout the galaxy. 

The previous survey of \hi\ absorption in the SMC by \citet{dic00} found a lower fraction of cold phase \hi\ in the SMC than in the Milky Way, which they suggested may be due to the lower abundance of coolants. Based on the theoretical model of the two-phase neutral interstellar medium (ISM) from \citet{wol95}, both the cold and warm phase can exist in pressure equilibrium in SMC-like conditions, but the range of pressures where this occurs is pushed to higher densities than in the Milky Way. \citet{dic00} also found that the spin temperatures of individual \hi\ absorption velocity components, which can be attributed to separate clouds of \hi, were generally colder than those found in the Milky Way. They suggested that the colder temperatures might be due to the fact that this \hi\ gas is photodissociated \htwo\ in cold clouds that would be molecular in the Milky Way. 

The current paper is the first in a series of three presenting the results from a new large Australia Telescope Compact Array (ATCA) survey looking for \hi\ absorption towards bright radio continuum sources across both of the Magellanic Clouds, which improves upon the previous survey due to the increased sensitivity and velocity resolution. Here we present the data and \hi\ temperature estimates from the SMC portion of the survey. Paper 2 (Liu et al. $in~prep$) will present the data and temperature estimates from the LMC portion. Paper 3 will discuss how the results in the Magellanic Clouds compare to the Milky Way and the Local Group, and identify how metallicity affects the conditions of the neutral atomic gas and the molecular-to-atomic transition. 

This paper is structured as follows: Section \ref{sec:obs} describes the design of the survey, the observations, and the methods of data reduction; Section \ref{sec:src_ex} details how we identified and extracted spectra for the continuum sources; the main results from the new survey, including \hi\ optical depths and average spin temperatures, are presented in Section \ref{sec:results}; Section \ref{sec:abs_comp} focuses on the identification and analysis of individual \hi\ absorption line features and includes a description of the method we use to measure the temperatures of individual \hi\ clouds associated with the line features; Section \ref{sec:discussion} compares the new survey  to previous work and to similar results from the Milky Way and LMC, provides an estimate of the cold gas fraction in the SMC, and discusses the apparent distribution of cold  \hi\ gas throughout the galaxy; the details of the conclusions we draw from this set of results are in Section \ref{sec:summary}.

\section{Observations} \label{sec:obs}

Observations were conducted with the Australia Telescope Compact Array (ATCA), between 2016 May 02 and 2018 Jan 14.  The ATCA is a six-element interferometer of 22m dishes with a maximum baseline of 6 km.  Five of the antennas can be relocated along a 3 km east-west rail track and the sixth antenna is at a fixed position 3 km west of the end of the rail track.  The array can be arranged into multiple configurations with extents ranging between 30 m and 6 km.  Our observations were conducted in the 6 km array configurations, which have the antennas distributed along the 3 km rail track for relatively uniform {\em u-v} coverage between baselines of $\sim 100$ m and 3000 m.  At 1420 MHz these array configurations return an angular resolution of approximately $10\arcsec$. 

We observed using the 1M-0.5k configuration Compact Array Broad-band Backend (CABB; \citealt{wil11}) correlator, which  simultaneously records 2048 channels across a 2 GHz bandwidth and 32 high spectral resolution zones (zooms) of 1 MHz, each with 2048 channels. The zooms were placed to give five concatenated 1 MHz zoom bands, overlapped by $50\%$ to obtain a flat bandpass, centred on 1419.5 MHz for \hi\, and at 1661.5, 1664.5, 1666.5 and 1719.5 MHz for all four OH lines. The velocity resolution in the \hi\ zoom band was $\sim 0.1$ km s$^{-1}$ and covered a total velocity range of $633$ km s$^{-1}$. 

We observed the standard ATCA primary calibrator PKS 1934-638 for 60 min per observing day for bandpass and amplitude calibration.  The unusually long bandpass observation was performed to ensure a very high signal-to-noise ($S/N \approx 550$) bandpass solution.  A phase calibrator source, 0252-712 or 0637-752, was observed for 2 min every 30 min  throughout an observing session for phase calibration.

The target sources were selected to simultaneously maximise optical depth sensitivity and coverage of a wide range of environmental conditions within the Magellanic Clouds. To achieve this we selected sources with flux density $S_{cont}>50$ mJy, coincident with areas of sufficient \hi\ column density (\Nhi\ $\gtrsim5\times10^{20}$ cm$^{-2}$), to detect \hi\ absorption by applying a cutoff in $S_{cont}$ vs \Nhi\ space. Our potential sources were chosen from an unpublished catalog of point sources in the LMC and SMC (private communication, Filipovic, 2016). The resolution of the catalogued SMC sources is limited to $\sim30\arcsec$. To capture sources unresolved in the catalogue but detectable with the $10\arcsec$ resolution of our array, we spent 12 hours of our allocated ATCA time to perform a quick continuum scan of all possible SMC sources. Our final source list had 29 sources in the SMC that are well distributed across the galaxy. The source density is such that there are multiple pairs of sources (12 pairs/triples) that are within the $80\%$ beam response radius and can be observed concurrently. We observed the pairs/triples together with a $50\%$ increase in observing time to accommodate reduced beam response for sources away from the pointing centre. A total of 21 fields were observed in the SMC.

We imaged and cleaned each observed field separately. The visibility data were first transformed into an image using the Miriad task {\sc invert}  with the image size set to $2050\times2050$ with a pixel size of $2\arcsec\times2\arcsec$ and Brigg's robustness parameter of $-2$. The data were imaged with a velocity of 0.2 km s$^{-1}$ per channel across the velocity range $70<\text{v}_{LSRK}<270$ km s$^{-1}$. We used {\sc invert} to produce both the imaged spectral line cube and a multifrequency synthesis (MFS) image, which we used as the continuum image. The typical RMS noise levels of the dirty spectral line cube and MFS image were $12$ mJy/beam and $0.6$ mJy/beam, respectively. We then used the Miriad task {\sc clean} to produce the images using a cut-off value of $\sim{2.5}\sigma$. As our observations mainly include point sources, we performed self-calibration to improve the calibration and reduce the sidelobe noise. With the cleaned MFS image as the model, we used the Miriad task {\sc selfcal} for components above 0.01 Jy/beam ($\sim10\sigma$). The Miriad task {\sc restor} was used to create the final maps from the clean components. For the three sets of fields that had significant overlap, we combined the two fields after imaging by creating a linear mosaic using the Miriad task {\sc linmos} .

\section{Source Spectra Extraction} \label{sec:src_ex}

\begin{figure}[t!]
\plotone{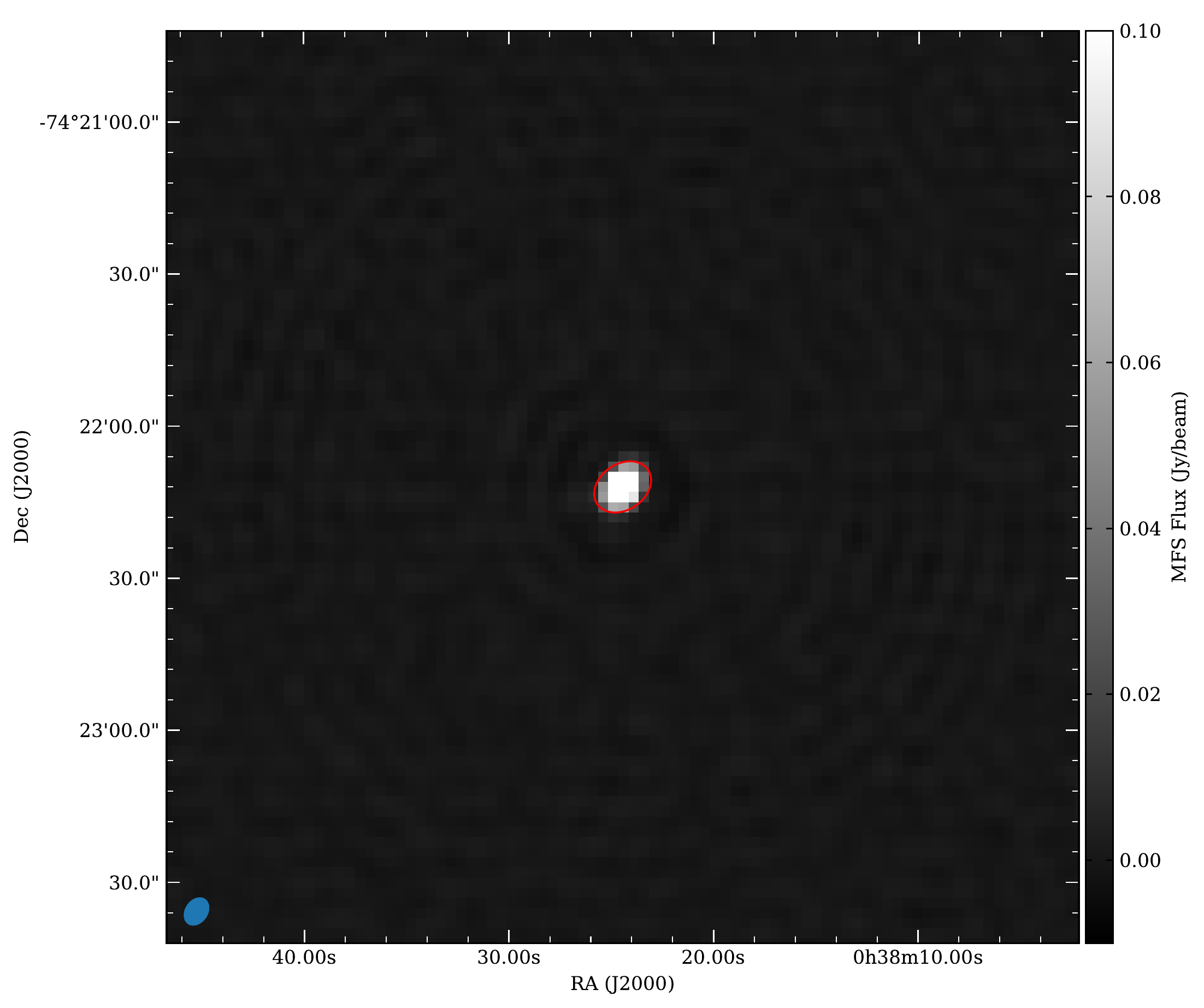}
\caption{Cut out of the multifrequency synthesis (MFS) image of field 0038-7422 centered on source 003824-742212, with the ellipse showing the source identified by Aegean. The complete figure set (55 images) is available in the online journal.
\label{fig:src_im}}
\end{figure}

While each field was centered on a known, bright continuum source, most fields also include fainter continuum sources. We used the python source-finding and extraction code Aegean \citep{han12,han18}, which was designed for radio data, on the multi-frequency synthesis (MFS) images for each field to identify all possible sources. Aegean classifies contiguous pixels above the noise threshold as islands and then fits each island with a collection of elliptical Gaussians, each of which are classified as components. We used the output ellipses, representing the FWHM of the fitted elliptical Gaussians, to define our sources.  To ensure that features in the spectrum of the continuum source would be detected with sufficient sensitivity, we only used sources detected at $S/N>10$ and with a peak flux $>20$ mJy in the MFS image (Figure Set 1 and Figure \ref{fig:src_im}). From the spectral cubes, we extracted all of the spectra for all pixels within the source ellipse and averaged them together. Each spectrum was weighted by the inverse of the square-root of the continuum level, which minimizes the noise \citep{dic92}, and produced spectra at a resolution of 0.2 km s$^{-1}$ per channel  (shown in gray in Figure \ref{fig:src_abs}). We also Hanning smoothed the high resolution spectra to produce a version at a resolution of 0.6 km s$^{-1}$ (using a Hanning function with width of 9 channels) and used this version for all of our analysis. The lower resolution is still sufficiently high to resolve narrow lines (FWHM $\sim{2}$ km s$^{-1}$), but decreases the noise in the spectra. 

\subsection{\hi\ Emission Spectra}

\begin{figure*}[t!]
\plotone{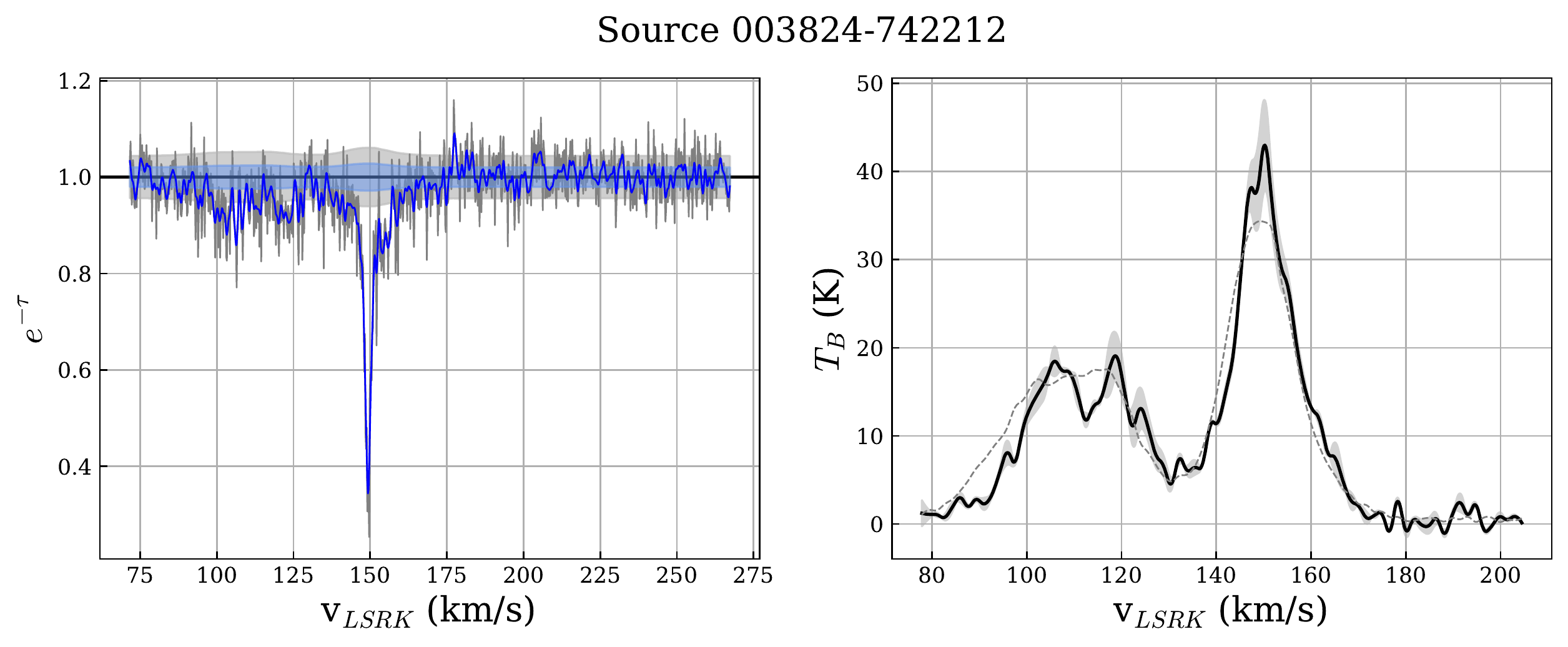}
\caption{ Extracted absorption and emission spectra, source 003824-742212. {\it Left:} Absorption. Lines show the absorption spectrum and shaded areas the  $1\sigma$ uncertainty level. Gray represents the highest spectral resolution of 0.2 km s$^{-1}$, blue represents the data smoothed to $\sim{0.6}$ km s$^{-1}$. {\it Right:}  Emission. The black line shows the extracted emission spectrum, the filled light gray area shows the estimated uncertainty. The dashed gray line shows the emission spectrum averaged over the primary beam size of individual ATCA dish ($34\arcmin$), which was used to scale the noise in the absorption spectrum (see Section \ref{sec:src_ex}). The complete figure set (55 images) is available in the online journal.
\label{fig:src_abs}}
\end{figure*}

We used the combined ATCA (interferometric) with Parkes (single-dish) data cube from \citet{sta99} to produce a spectrum of the \hi\ emission for each source. The cube has a resolution of $98\arcsec$ (30 pc) and velocity resolution of 1.65 km s$^{-1}$. We took the mean of all spectra found within the beam concentric with the identified continuum source. We interpolated the spectrum using a quadratic spline to match the channel size of the absorption spectrum of $\Delta\text{v}\sim0.2$ km s$^{-1}$. The interpolation is only accurate under the assumption that the \hi\ emission profiles are smoothly varying. This assumption is likely to be reasonable given the observed wide \hi\ emission features (FWHM $\gtrsim{20}$ km s$^{-1}$) and that even narrow $\sim{5}$ km s$^{-1}$ emission lines would be resolved with 1.65 km s$^{-1}$ channels and be observable in the original spectra. The one exception would be the possibility of narrow self-absorption from intervening cold \hi\ along the line of sight. However, when simulating the effects of even low optical depth ($\tau\sim0.1$) self-absorption on a broader emission line, the self-absorption would still be observable in the \hi\ emission spectra. Figure \ref{fig:src_abs} shows an extracted emission spectrum in which the shaded gray area indicates the the estimated uncertainty. Future \hi\ emission maps of the SMC from the Australian Square Kilometre Array Pathfinder (ASKAP) will provide emission spectra with higher spatial and spectral resolution.

\subsection{Noise Spectra} \label{subsec:abs_noise}

For the emission spectra, we measured the off-line RMS using one of two velocity ranges: either $\sim{80}<\textrm{v}_{LSRK}<100$ km s$^{-1}$ or $\sim{180}<\textrm{v}_{LSRK}<200$ km s$^{-1}$. We accounted for the additional uncertainty due to the variation in the emission spectra found within one beam centered on the continuum source by scaling the off-line RMS by the $1\sigma$ variation in the emission spectra. This is shown by the grey filled region on the emission spectrum in Figure \ref{fig:src_abs}. 

Although the interferometric ATCA observations filter out the large-scale \hi\ emission of the SMC, the emission nonetheless has an impact on the noise in the spectra. It will increase the system temperature of individual ATCA antennas as a function of velocity. To estimate the noise spectra, $\sigma_{\tau}(\text{v})$, we first computed the RMS in the absorption spectra away from the SMC \hi\ emission over the velocity range $225<\text{v}_{LSRK}<270$ km s$^{-1}$ and then scaled it by  $\left((T_{B}(\text{v})+T_{sys})/T_{sys}\right)$, where $T_{B}(\text{v})$ is the average brightness temperature spectrum in the beam of an individual ATCA antenna (shown by the dashed grey lines in Figure \ref{fig:src_abs}) and $T_{sys}$ is the system temperature of the receiver. This takes into account the effective increase to the brightness temperature seen by the receiver due to the large-scale \hi\ emission. We set $T_{sys}=44.7$ K, the expected system temperature of the receivers for observations at high elevation with good weather\footnote{http://www.narrabri.atnf.csiro.au/myatca/interactive\_senscalc.html}. The actual system temperature will vary during and between observations, but by choosing the lowest reasonable value we ensure that we are not underestimating the noise contribution from the brightness temperature in the primary beam. To produce the final noise spectrum, we scaled the off-line RMS value by the average of the \hi\ emission spectra from the ATCA+Parkes map across the primary beam of the ATCA at 21 cm ($34\arcmin$) multiplied by the ATCA antenna efficiency of $0.50$. 

\section{\hi\ Absorption Properties} \label{sec:results}

We have extracted absorption spectra for a total of 55 sources across the SMC. Figures \ref{fig:src_im} and \ref{fig:src_abs} show the MFS image and extracted absorption spectrum and emission spectra respectively for an example source (0038-7422). Each absorption spectrum is presented in terms of opacity ($e^{-\tau}$), which is calculated using the average continuum flux in the off-line velocities of the spectra $\sim{225}<\textrm{v}_{LSRK}<270$ km s$^{-1}$:
\begin{equation}
e^{-\tau(\text{v})} = S(\text{v})/S_{cont}.
\label{eqn:opacity}
\end{equation}
The locations, continuum source fluxes, peak optical depth, $\tau_{peak}$, and average $1\sigma$ uncertainty on $\tau$, $\sigma_{\tau}$, measured off-line for the smoothed spectra (velocity resolution 0.6 km s$^{-1}$), are presented in Table \ref{table:srcs}. For each absorption spectrum we calculate the integral over the entire image velocity range ($\sim{70}<\textrm{v}_{LSRK}<270$ km s$^{-1}$) giving us the equivalent width (EW) of the absorption:
\begin{equation}
\text{EW}=\int (1-e^{-\tau(\text{v})})d\text{v}.
\label{eqn:EW}
\end{equation}
From the emission, we calculate the column density uncorrected for optical depth using the integral of the spectrum over the whole velocity range covered \text{by the existing observations} ($\sim{80}<\textrm{v}_{LSRK}<200$ km s$^{-1}$):
\begin{equation}
N_{\text{H,uncor}} = 1.823\times10^{18} \int T_{b}(\text{v})d\text{v}~\text{cm}^{-2}.
\label{eq:n_uncor}
\end{equation}
The values for equivalent width and $N_{\text{H,uncor}}$ are listed in columns 8 and 9 of Table \ref{table:srcs}.

We use the equivalent width measurements to determine whether or not absorption is detected towards each continuum source. If the equivalent width is measured at $>3\sigma$, we classify that as a detection of absorption. This method avoids any assumptions about the shape of the absorption features and allows us to detect the low optical depth, potentially wide features, that we see in some of the absorption spectra (e.g. in the source shown in Figure \ref{fig:src_abs}). The use of the equivalent width avoids spurious or marginal absorption detections that can occur when only considering the peak in each absorption spectrum. While a $3\sigma$ threshold may seem low, we have visually inspected all spectra with identified absorption and find it to be reasonable. Some of the lowest signal-to-noise detections of absorption ($\sim3\sigma$), based on the equivalent width measurement, have narrow absorption lines where the peak optical depth is detected at $\gtrsim5\sigma$ (e.g., source 013032-731740).

Using this method, we detect absorption in 37 out of 55 continuum sources with $S/N>10$ and a peak flux density $>20$ mJy. Amongst the 18 spectra in which no absorption was detected, many have low sensitivity and weak to moderate optical depths: 10 sources have spectra with $\sigma_{\tau}>0.5$, and the remaining 8 are all found in regions with low \hi\ column densities and outside the main body of the galaxy (i.e. outside the region bounded by $N_{\text{H}}\leq2\times10^{21}$ cm$^{-2}$, the contour level shown in Figure \ref{fig:tau_map}). Given the low sensitivity of a significant number of sources, our detection rate of $67\%$ may be an underestimate of the true fraction of lines-of-sight with cold \hi. If we only consider sources with good $\tau$ sensitivity of $\sigma_{\tau}\leq0.2$ (33 sources), then we have an absorption detection rate of $88\%$, and it is likely that if we improved the sensitivity of some of the observations we would detect lower optical depth \hi\ absorption.

\startlongtable
\begin{deluxetable*}{ccccccccccccc}
\tablecaption{Absorption Spectra Quantities \label{table:srcs}}
\tabletypesize{\footnotesize}
\tablehead{
\colhead{Field} & \colhead{Source Name} & \colhead{R.A.} & \colhead{Dec.} & \colhead{$S_{cont}$} & \colhead{$\tau_{peak}$\tablenotemark{a}} & \colhead{$\sigma_{\tau}$\tablenotemark{a}} & \colhead{EW} & \colhead{$N_{\text{H,uncor}}$} &  \colhead{$f_{\text{H,cor,iso}}$} & \colhead{$<$T$_{s}>$}  \\
\colhead{} & \colhead{} & \colhead{(J2000)} & \colhead{(J2000)} & \colhead{(mJy)} & \colhead{} & \colhead{} & \colhead{(km s$^{-1}$)} & \colhead{($10^{21}$ cm$^{-2}$)} & \colhead{} & \colhead{(K)}
} 
\startdata
0026-7355 & 002906-735333 & 7.2771 & -73.8925 & 90.83 & 1.1 & 0.10 & 7.4$\pm$0.6 & 1.3$\pm$0.3 & 1.09 & 96.3$\pm$24.0 \\
0038-7350 & 003809-735024 & 9.5387 & -73.8401 & 109.35 & 0.8 & 0.04 & 5.1$\pm$0.3 & 2.7$\pm$0.2 & 1.04 & 292.8$\pm$28.1 \\
0038-7422 & 003824-742212 & 9.6018 & -74.3700 & 203.35 & 1.1 & 0.02 & 5.8$\pm$0.1 & 2.2$\pm$0.2 & 1.07 & 206.6$\pm$20.3 \\
0040-7300 & 004229-730406 & 10.6216 & -73.0685 & 77.10 & 4.0 & 0.18 & 31.6$\pm$1.2 & 4.8$\pm$0.3 & 1.34 & 83.2$\pm$ 6.6 \\
0040-7300 & 003800-725210 & 9.5040 & -72.8697 & 41.37 & 2.4 & 0.24 & 3.9$\pm$1.4 & 1.6$\pm$0.2 & 1.06 & $>$ 215.9 \\
0040-7300 & 003754-725156 & 9.4775 & -72.8657 & 101.36 & 0.4 & 0.09 & 6.7$\pm$0.6 & 1.6$\pm$0.2 & 1.01 & 128.7$\pm$19.1 \\
0041-7146 & 003947-713735 & 9.9486 & -71.6265 & 49.75 & 0.4 & 0.08 & 0.8$\pm$0.5 & 0.4$\pm$0.1 & 1.00 & $>$ 146.6 \\
0041-7146 & 004047-714559 & 10.1999 & -71.7665 & 354.93 & 0.04 & 0.01 & 0.4$\pm$0.1 & 0.5$\pm$0.1 & 1.00 & 805.7$\pm$266.1 \\
0041-7146 & 003939-714141 & 9.9155 & -71.6949 & 55.60 & 0.3 & 0.07 & -1.7$\pm$0.4 & 0.6$\pm$0.1 & 1.00 & $>$ 247.4 \\
0048-7412 & 004808-741205 & 12.0350 & -74.2017 & 50.95 & 0.6 & 0.10 & 10.8$\pm$0.6 & 1.9$\pm$0.2 & 1.04 & 95.9$\pm$13.6 \\
0053-7313 & 005238-731245 & 13.1592 & -73.2126 & 63.71 & 6.7 & 0.09 & 25.0$\pm$0.4 & 9.1$\pm$0.6 & 1.30 & 200.8$\pm$12.8 \\
0054-7227 & 005218-722708 & 13.0784 & -72.4524 & 231.96 & 0.6 & 0.06 & 6.4$\pm$0.4 & 3.2$\pm$0.4 & 1.06 & 271.8$\pm$42.0 \\
0054-7227 & 005519-721050 & 13.8291 & -72.1808 & 28.63 & 3.6 & 0.31 & 12.8$\pm$1.8 & 2.4$\pm$0.4 & 1.08 & 105.1$\pm$22.1 \\
0054-7227 & 004718-723948 & 11.8288 & -72.6636 & 57.28 & 7.3 & 1.28 & -8.8$\pm$5.8 & 3.5$\pm$0.3 & 1.00 & $>$ 112.2 \\
0054-7227 & 004956-723554 & 12.4869 & -72.5986 & 150.76 & 3.1 & 0.13 & 13.2$\pm$0.8 & 4.2$\pm$0.4 & 1.16 & 173.4$\pm$18.2 \\
0054-7227 & 005557-722606 & 13.9877 & -72.4352 & 52.42 & 5.1 & 0.53 & -1.8$\pm$2.7 & 6.2$\pm$0.6 & 1.04 & $>$ 414.9 \\
0054-7227 & 005555-722556 & 13.9792 & -72.4324 & 63.62 & 3.2 & 0.15 & 14.9$\pm$0.9 & 6.3$\pm$0.6 & 1.16 & 231.4$\pm$26.5 \\
0054-7227 & 005337-723144 & 13.4074 & -72.5291 & 90.76 & 2.4 & 0.16 & 14.5$\pm$1.0 & 4.9$\pm$0.4 & 1.13 & 185.0$\pm$20.5 \\
0056-7107 & 005652-712300 & 14.2195 & -71.3835 & 44.87 & 0.7 & 0.10 & 8.9$\pm$0.6 & 1.5$\pm$0.1 & 1.07 & 92.8$\pm$10.7 \\
0056-7107 & 010029-713826 & 15.1240 & -71.6408 & 20.07 & 3.5 & 0.22 & 12.9$\pm$1.3 & 3.7$\pm$0.3 & 1.32 & 158.8$\pm$20.1 \\
0056-7107 & 005611-710707 & 14.0478 & -71.1188 & 188.96 & 0.2 & 0.02 & 2.4$\pm$0.1 & 1.0$\pm$0.1 & 1.02 & 236.4$\pm$34.8 \\
0056-7107 & 005820-713040 & 14.5853 & -71.5114 & 16.18 & 2.8 & 0.32 & 1.2$\pm$1.7 & 1.7$\pm$0.3 & 1.11 & $>$ 189.8 \\
0058-7413 & 005732-741243 & 14.3855 & -74.2120 & 339.90 & 0.7 & 0.01 & 1.5$\pm$0.1 & 0.7$\pm$0.1 & 1.00 & 239.3$\pm$52.7 \\
0058-7413 & 005636-740315 & 14.1538 & -74.0543 & 19.05 & 1.3 & 0.18 & 4.5$\pm$1.1 & 1.3$\pm$0.2 & 1.03 & 162.0$\pm$49.7 \\
0101-7138 & 005820-713040 & 14.5859 & -71.5111 & 19.82 & 3.7 & 0.23 & -5.9$\pm$1.4 & 1.7$\pm$0.3 & 1.11 & $>$ 227.7 \\
0101-7138 & 010029-713826 & 15.1245 & -71.6406 & 100.52 & 6.1 & 0.05 & 10.5$\pm$0.3 & 3.7$\pm$0.3 & 1.36 & 194.5$\pm$15.8 \\
0110-7135 & 010930-713456 & 17.3785 & -71.5823 & 72.29 & 0.9 & 0.12 & 14.7$\pm$0.8 & 2.0$\pm$0.3 & 1.06 & 74.1$\pm$12.3 \\
0110-7135 & 010932-713452 & 17.3842 & -71.5814 & 29.73 & 1.1 & 0.17 & 12.2$\pm$1.0 & 2.0$\pm$0.3 & 1.05 & 88.5$\pm$14.5 \\
0110-7227 & 011005-722647 & 17.5226 & -72.4467 & 121.46 & 0.6 & 0.03 & 13.1$\pm$0.2 & 3.5$\pm$0.4 & 1.09 & 144.0$\pm$16.5 \\
0110-7227 & 011035-722807 & 17.6491 & -72.4687 & 33.67 & 2.9 & 0.12 & 31.8$\pm$0.7 & 3.7$\pm$0.4 & 1.36 & 64.6$\pm$ 6.9 \\
0111-7314 & 011432-732143 & 18.6369 & -73.3622 & 39.67 & 6.9 & 0.23 & 31.8$\pm$1.4 & 6.0$\pm$0.6 & 1.38 & 104.2$\pm$10.6 \\
0111-7314 & 011049-731428 & 17.7069 & -73.2412 & 321.62 & 0.7 & 0.03 & 10.8$\pm$0.2 & 5.0$\pm$0.5 & 1.11 & 252.3$\pm$23.4 \\
0111-7314 & 011056-731404 & 17.7369 & -73.2346 & 44.05 & 1.6 & 0.11 & 22.4$\pm$0.7 & 5.2$\pm$0.4 & 1.14 & 126.3$\pm$11.2 \\
0111-7314 & 011047-731400 & 17.6989 & -73.2334 & 27.35 & 1.7 & 0.14 & 26.3$\pm$0.9 & 5.1$\pm$0.5 & 1.15 & 106.7$\pm$11.7 \\
0111-7314 & 011132-730209 & 17.8860 & -73.0361 & 66.30 & 2.2 & 0.11 & 22.8$\pm$0.7 & 5.6$\pm$0.6 & 1.30 & 135.1$\pm$14.1 \\
0111-7314 & 010919-725600 & 17.3314 & -72.9334 & 21.14 & 3.8 & 0.35 & 1.1$\pm$2.0 & 5.1$\pm$0.4 & 1.08 & $>$ 476.0 \\
0115-7322 & 011628-731438 & 19.1207 & -73.2441 & 51.98 & 3.9 & 0.07 & 14.9$\pm$0.4 & 5.4$\pm$0.4 & 1.31 & 197.7$\pm$14.5 \\
0115-7322 & 011432-732142 & 18.6364 & -73.3619 & 99.80 & 6.7 & 0.04 & 17.1$\pm$0.2 & 6.0$\pm$0.6 & 1.53 & 193.6$\pm$17.9 \\
0115-7322 & 011049-731427 & 17.7067 & -73.2410 & 83.32 & 0.6 & 0.06 & 5.9$\pm$0.4 & 5.0$\pm$0.5 & 1.07 & 463.4$\pm$51.5 \\
0119-7106 & 011919-710523 & 19.8317 & -71.0897 & 51.53 & 1.4 & 0.26 & 14.9$\pm$1.5 & 1.1$\pm$0.2 & 1.04 & 39.5$\pm$ 8.8 \\
0119-7106 & 011917-710537 & 19.8237 & -71.0939 & 70.04 & 2.0 & 0.19 & 0.5$\pm$1.1 & 1.1$\pm$0.2 & 1.04 & $>$ 184.3 \\
0124-7351 & 012348-735033 & 20.9518 & -73.8427 & 31.20 & 3.9 & 0.28 & 22.2$\pm$1.8 & 2.0$\pm$0.3 & 1.15 & 49.7$\pm$ 8.5 \\
0124-7351 & 012350-735042 & 20.9587 & -73.8451 & 34.41 & 1.5 & 0.25 & 29.5$\pm$1.5 & 1.9$\pm$0.3 & 1.08 & 36.0$\pm$ 5.3 \\
0124-7351 & 012323-735606 & 20.8486 & -73.9352 & 28.41 & 2.4 & 0.26 & 30.9$\pm$1.6 & 1.5$\pm$0.2 & 1.08 & 26.6$\pm$ 4.2 \\
0130-7333 & 012930-733310 & 22.3758 & -73.5530 & 552.03 & 0.2 & 0.06 & -1.3$\pm$0.4 & 1.5$\pm$0.4 & 1.00 & $>$ 779.0 \\
0130-7333 & 013032-731740 & 22.6345 & -73.2946 & 41.07 & 7.7 & 0.78 & 17.6$\pm$5.8 & 2.1$\pm$0.4 & 1.05 & 65.8$\pm$25.5 \\
0130-7333 & 013228-734123 & 23.1205 & -73.6899 & 41.83 & 4.4 & 0.79 & 10.0$\pm$4.4 & 1.0$\pm$0.5 & 1.06 & $>$ 41.3 \\
0130-7333 & 013215-733902 & 23.0637 & -73.6508 & 42.35 & 3.9 & 0.81 & -25.6$\pm$3.9 & 1.1$\pm$0.6 & 1.02 & $>$ 52.7 \\
0130-7333 & 012931-733318 & 22.3799 & -73.5551 & 34.46 & 5.1 & 0.51 & -0.6$\pm$2.5 & 1.6$\pm$0.4 & 1.11 & $>$ 119.8 \\
0130-7333 & 013213-733905 & 23.0569 & -73.6515 & 17.97 & 5.5 & 1.07 & -23.2$\pm$6.6 & 1.1$\pm$0.6 & 1.02 & $>$ 31.3 \\
0130-7333 & 012629-732714 & 21.6222 & -73.4540 & 70.79 & 5.8 & 0.54 & 3.7$\pm$2.3 & 3.3$\pm$0.4 & 1.23 & $>$ 263.4 \\
0130-7333 & 012639-731501 & 21.6666 & -73.2505 & 48.08 & 9.4 & 0.51 & 20.6$\pm$2.8 & 2.9$\pm$0.4 & 1.06 & 77.8$\pm$15.8 \\
0130-7333 & 013147-734942 & 22.9482 & -73.8283 & 19.21 & 5.1 & 1.06 & 55.4$\pm$7.4 & 1.8$\pm$0.3 & 1.03 & 17.5$\pm$ 4.2 \\
0130-7333 & 012924-733152 & 22.3506 & -73.5311 & 40.90 & 6.0 & 1.00 & 3.9$\pm$6.3 & 1.3$\pm$0.3 & 1.02 & $>$ 37.1 \\
0130-7333 & 013243-734413 & 23.1822 & -73.7371 & 33.90 & 4.9 & 0.65 & 11.5$\pm$3.7 & 1.2$\pm$0.5 & 1.09 & 55.0$\pm$31.7 \\
\enddata
\tablenotetext{a}{in the spectra Hanning smoothed to $0.6$ km s$^{-1}$ spectral resolution with $0.2$ km s$^{-1}$ channels}
\end{deluxetable*}

\subsection{Optical Depth Correction to the \hi\ Column Density \label{subsec:N_cor}}  

The \hi\ absorption line spectra provide the only direct means of determining the optical depth of the \hi\ along the line of sight. With \hi\ emission spectra alone, the column density is typically determined by assuming optically thin emission (Equation \ref{eq:n_uncor}). If the gas is not optically thin, this will result in an underestimate of the \hi\ column density. We correct for optical depth using our absorption spectra on a channel-by-channel basis using
\begin{equation}
N_{\text{H,cor,iso}} = 1.823\times10^{18} \int \frac{T_{b}(\text{v})\times{\tau}(\text{v})}{1-e^{-\tau(\text{v})}}d\text{v}~\text{cm}^{-2}
\end{equation}
assuming the \hi\ is isothermal at a given velocity \citep{dic82}. In Table \ref{table:srcs}, we report the correction factor to the column density for the isothermal case ($f_{\text{H,cor,iso}}$) where $f_{\text{H,cor,iso}}=N_{\text{H,cor,iso}}/N_{\text{H,uncor}}$. While there are likely to be multiple parcels of gas with different spin temperatures at the same velocity along many lines-of-sight, \citet{kim14} show that for three-dimensional hydrodynamical simulations of the ISM the isothermal approximation is within $5\%$ of the true \hi\ column density for $90\%$ of the lines-of-sight. Here, we are interested in an estimate of the total column density of \hi, but we do not rely heavily on these values. The corrections factors are typically small:  an average of $f_{\text{H,cor,iso}}\sim{1.1}$, which is similar to the optical depth corrections from \citet{sta99}, and indicates that there is not a large amount of unaccounted for \hi\ mass due to optically thick emission in the SMC. We note that we do not observe any flattening in the emission spectra where we see high optical depth, which indicates that the method used by \citet{bra09} and \citet{bra12} to estimate \hi\ optical depth from emission spectra is not accurate for the SMC. We defer a more thorough exploration of optical depth corrections on the column density to a forthcoming paper.

\subsection{The Optical Depth of \hi\ in the SMC} \label{subsec:tau} 

\begin{figure}[t!]
\epsscale{1.2}
\plotone{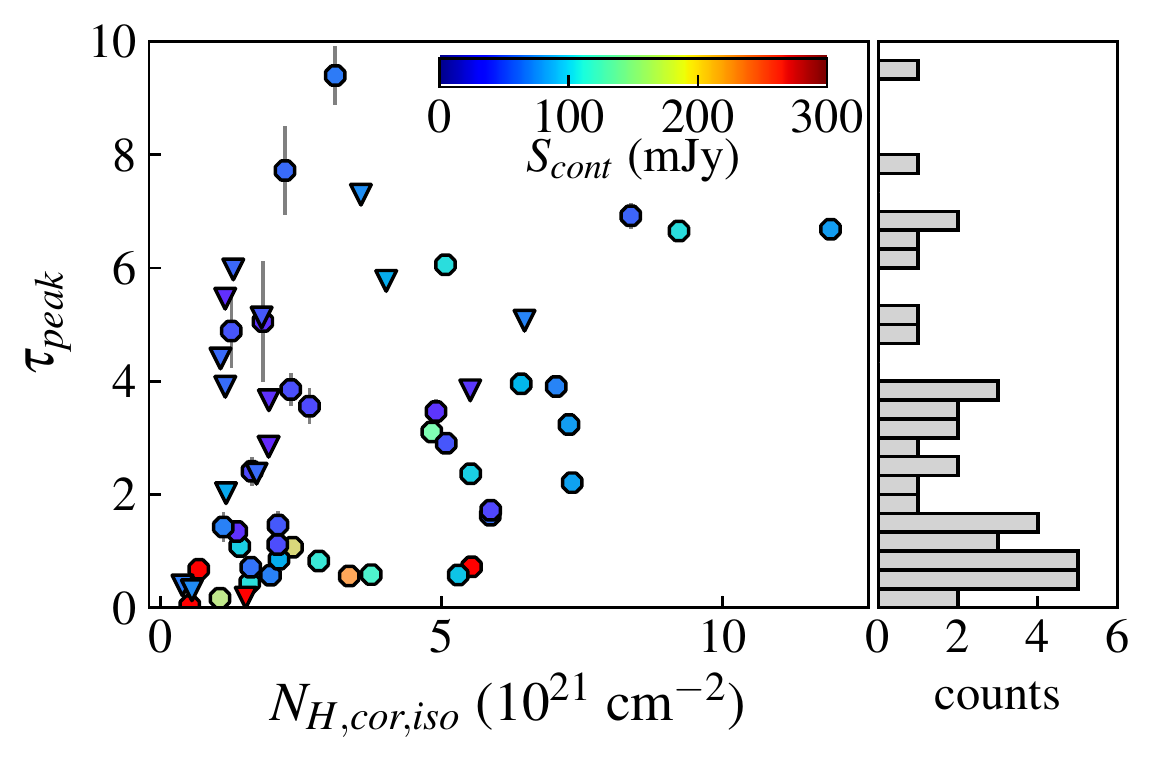}
\caption{Observed peak \hi\ absorption optical depth ($\tau_{peak}$) vs.\ optical depth corrected \hi\ column density ($N_{H,corr,iso}$) 
{\it Left:} Continuum sources are represented by circles (absorption detected) or triangles (no detected absorption) and color coded according to the strength of their flux ($S_{cont}$) in mJy (color scale bar). {\it Right:} Histogram showing the number of sources per band of optical depth.)
\label{fig:tau_peak}}
\end{figure}

For spectra with detected absorption we observe $0.2<\tau_{peak}<9.4$. Of the 37 sources with detected absorption, 25 of those ($67\%$) show optically thick \hi\ gas with $\tau_{peak}\ge1$. Figure \ref{fig:tau_peak} shows the full distribution of the peak $\tau$ values and demonstrates that there is no clear correlation between the peak optical depth and the total column density of \hi\ gas; the presence of high optical depth \hi\ is not dependent on the total amount of gas along the line-of-sight and there is optically thick \hi\ gas at low column densities. The strength of the continuum source (see color scale, Figure \ref{fig:tau_peak}) largely determines the noise level in the absorption spectrum, but there is no indication that the detection of absorption is biased by signal strength. This implies that the noise levels in many of the weaker sources would permit the detection of moderate to low optical depth absorption ($\tau<1$) by \hi\ gas along the line of sight, if it were present. 

\begin{figure}[t!]
\epsscale{1.2}
\plotone{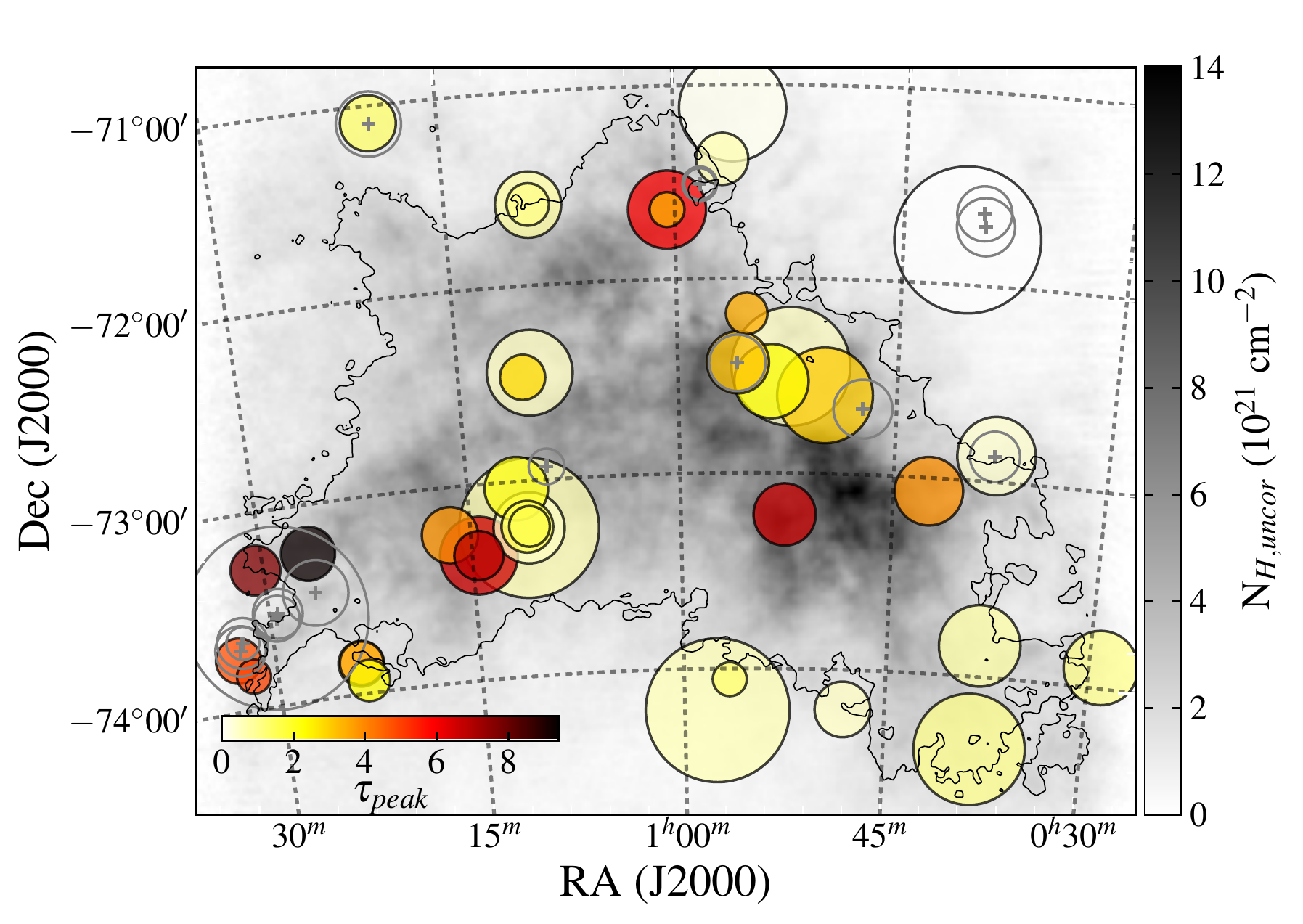}
\caption{Map of detected continuum sources. Sources with detected absorption ($\tau_{peak}>3\sigma_{\tau}$) are represented by colored circles, where the color indicates peak absorption ($\tau_{peak}$, color scale, bottom left), and circle size represents the source strength. Sources with no detected absorption are shown as grey circles with crosses. The background gray scale image shows the column density of \hi\ (assuming optically thin emission), and the main body of the galaxy is indicated by the black contour line, representing $N_{\text{H,uncor}}=2\times10^{21}$ cm$^{-2}$.
\label{fig:tau_map}}
\end{figure}

Figure \ref{fig:tau_map} shows that the sources with detected absorption are found across the SMC, and that most of the optically thick \hi\ is located within the main body of the galaxy. The peak optical depth does not show any strong systematic regional trends and appears to be randomly distributed. The lower $\tau_{peak}$ sources tend to be located on the edge of or outside the main body of the SMC. Sources with no detected absorption are also either outside the main body of the SMC, or have high noise due to the low continuum strength of the source. There is also small scale variation in the \hi\ optical depth along the line of sight (e.g., the two sources at similar positions in the North with $\tau_{peak}\sim4$ and $\tau_{peak}\sim6$). 

\subsection{Average Spin Temperatures of the \hi\ Gas} \label{subsec:Ts} 

\begin{figure}[t!]
\epsscale{1.1}
\plotone{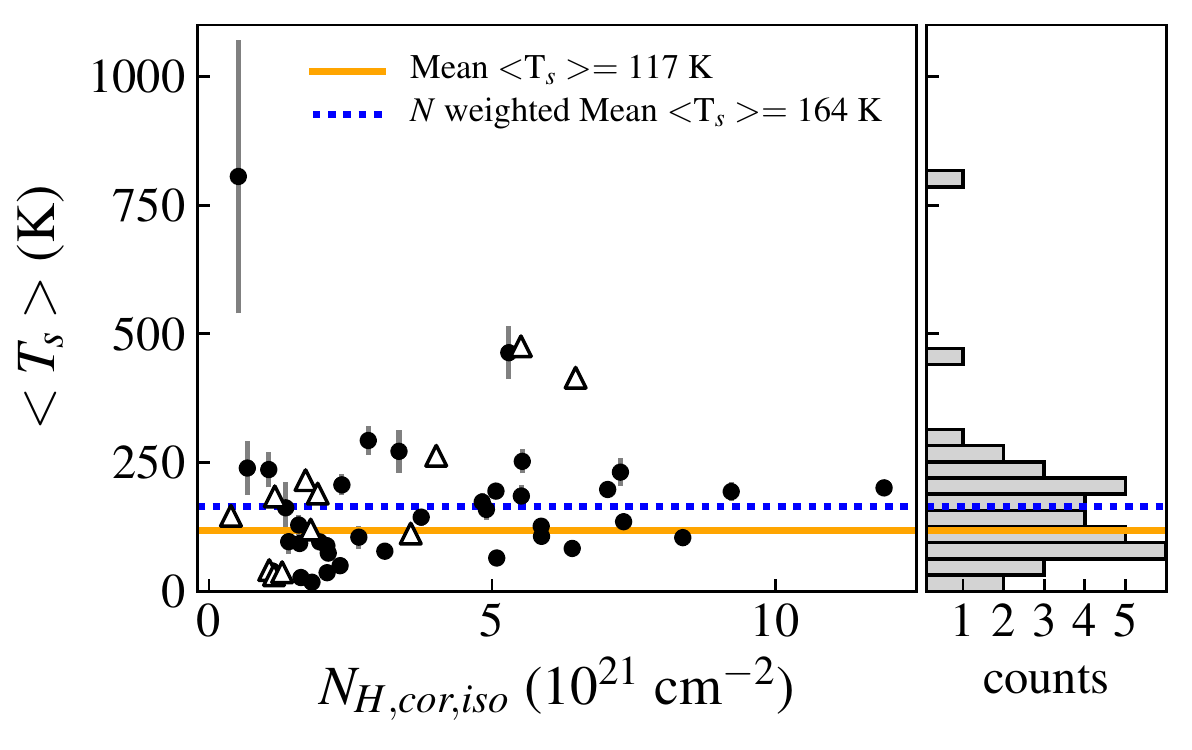}
\caption{Average \hi\ spin temperature ($<T_{s}>$) vs. optical depth corrected column density ($N_{\text{H,cor,iso}}$). {\it Left:} Filled circles show average \hi\ spin temperatures ($<T_{s}>$) along lines of sight to sources for which $<T_{s}>$ are detected at $>3\sigma$, and open triangles represent those for which upper limits are measured with $<3\sigma$. The blue dashed line shows the column density ($N$) weighted mean of $<T_{s}>$ and the orange line shows the arithmetic mean of $<T_{s}>$, which were measured to be $150.8\pm12.2$ K and $117.2\pm101.7$, respectively. These values were calculated as described in Section 4.3.
{\it Right:} Histogram shows the number of sources per $\sim{30}$ K  temperature band for which $<T_{s}>$ is detected at $>3\sigma$.
\label{fig:Ts_avg}}
\end{figure}

The excitation temperature of the \hi\ gas is often referred to as the spin temperature ($T_{s}$) as the 21 cm line is a spin-flip transition, and is determined by the Boltzmann relationship between the populations of the upper and lower hyperfine energy states of the ground level of hydrogen. At the densities of the cold neutral medium, collisions can thermalize the 21 cm transition so that the spin temperature is identical to the kinetic temperature ($T_{k}$) of the gas \citep{fie58}. Measuring the spin temperature requires both absorption and emission observations:
\begin{equation}
T_{s}(\text{v}) = \frac{T_{B}(\text{v})}{1-e^{-\tau(\text{v})}}.
\label{eqn:Ts}
\end{equation}
For sources where we detect absorption, we can use the integrated emission ($N_{\text{H,uncor}}$) and absorption (EW) to measure the line-of-sight average spin temperature, $<T_{s}>$:
\begin{equation}
<T_{s}> = \frac{\int T_{B}(\textrm{v})d\text{v}}{\int 1-e^{-\tau(\text{v})}d\text{v}}\sim \left<\frac{n(s)}{[n(s)/T_{s}(s)]}\right>,
\label{eqn:Ts_avg}
\end{equation}
where $n(s)$ is the volume density along the line-of-sight (for a more detailed derivation see \citealt{dic00}). From this we see that $<T_{s}>$ is the volume density weighted harmonic mean of $T_{s}(s)$. In physical terms, the average spin temperature represents a weighted average of the WNM and CNM temperatures along the line of sight. Our measured values of $<T_{s}>$ are listed in Table \ref{table:srcs} with upper limits for sources where no absorption is detected above the $3\sigma$ level in the equivalent width measurement. We measure average spin temperatures that range from very low values of $<T_{s}>\sim{30}$ K to as high as $<T_{s}>\sim{800}$ K (Figure \ref{fig:Ts_avg}). The observed distribution of $<T_{s}>$ has an arithmetic mean and standard deviation of $<T_{s}>=117.2\pm101.7$ K, which was calculated including the lower limits using the cenfit routine in the R package NADA \citep{lee17,hel05}. 

We also sum all of the column densities and equivalent widths separately and then take the ratio:
\begin{equation}
\frac{\sum N_{\text{H,uncor}}}{\sum EW}=\frac{\int n(s)ds}{C_{0}^{-1}\int[n(s)/T_{s}(s)]ds}
\end{equation}
\textbf{where $C_{0}=1.823\times10^{18}$ cm$^{-2}/(\textrm{K km s}^{-1})$}, which results in a column density weighted harmonic mean of $T_{s}(s)$. We find a column density weighted mean $<T_{s}>$ of $150.8\pm12.2$ K where the uncertainty is the propagated error. If we make a sensitivity cut of $\sigma_{\tau}<0.15$ and take the high $S/N$ observations (27 out of the total 55 sources), then we measure a column density weighted mean $<T_{s}>$ of $164.1\pm14.4$ K, which is consistent with the value for the entire sample.

Because $<T_{s}>$ measurements require the detection of absorption, these values tell us about the relative temperatures and mass fractions of the CNM and WNM when both are detected along the line-of-sight. Assuming that all of the optical depth is coming from the cold component and the average cloud temperature is a representative value of the cold phase temperature, then the equivalent width is
\begin{equation}
EW = \frac{N_{c}}{C_{0}T_{c}}
\end{equation}
where $N_{c}$ is the column density of the cold phase gas. The fraction of gas in the cold phase is
\begin{equation}
f_{c}=\frac{N_{c}}{N_{w}+N_{c}}
\end{equation}
where $N_{w}$ is the column density of the warm phase gas. Given that most of the optical depth corrections to the column density are small (see Section \ref{subsec:N_cor}), then $N_{unc}\simeq{N_{w}+N_{c}}$ so that
\begin{equation}
N_{c}\simeq{f_{c}N_{unc}}.
\end{equation}
This then results in $<T_{s}>$ in terms of $f_{c}$ and $T_{c}$:
\begin{equation}
\label{eqn:fc}
<T_{s}>=\frac{N_{unc}}{EW}\simeq \frac{T_{c}}{f_{c}}.
\end{equation}
 The observed low $<T_{s}>$ values of $30-50$ K suggest that the CNM component of the \hi\ is indeed cold, at least $30-50$ K, and can be a significant fraction of the gas along some sight lines. Observationally, we are biased towards colder temperatures that show larger absorption.  

\begin{figure}[t!]
\epsscale{1.2}
\plotone{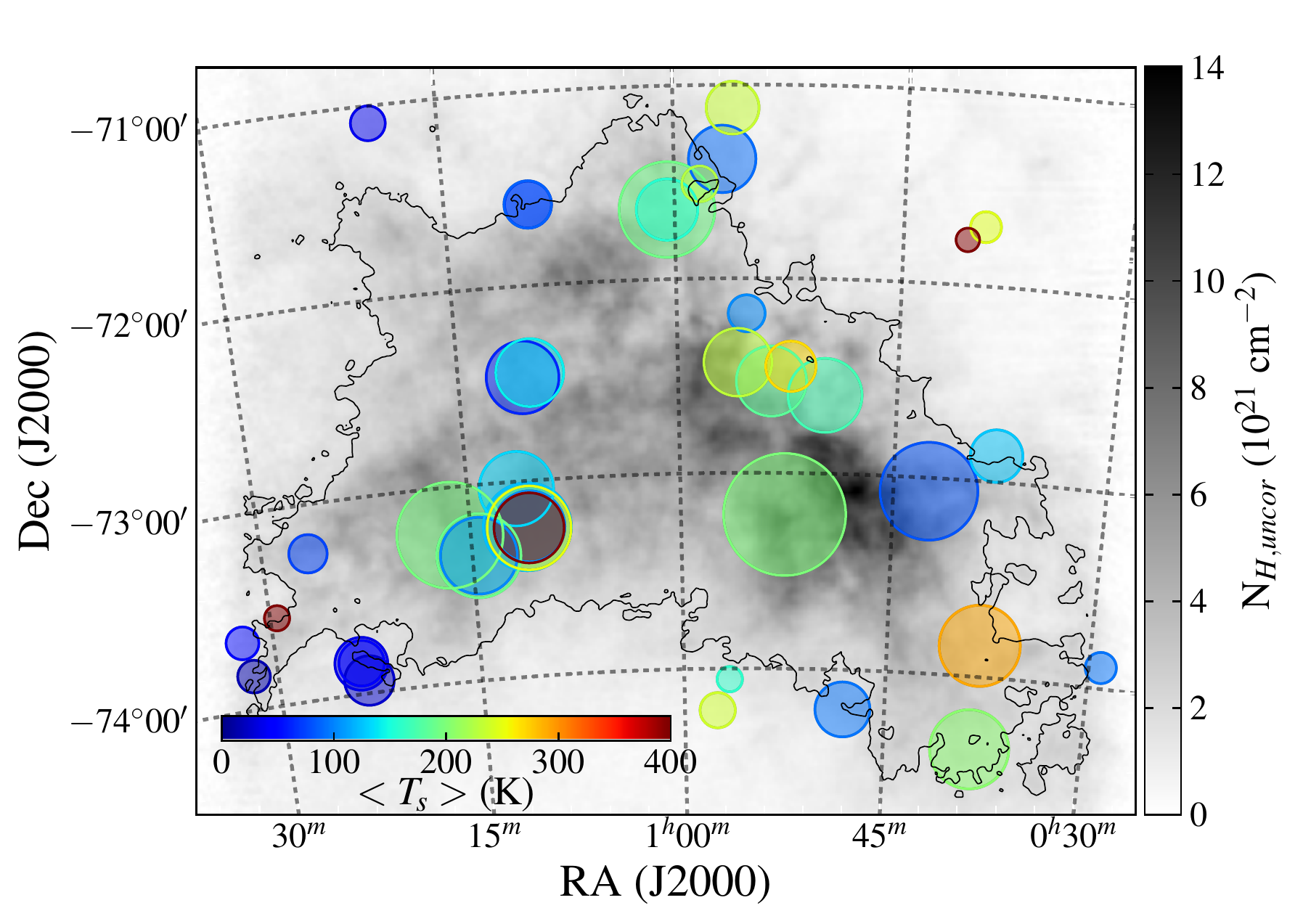}
\caption{Spatial distribution of measured spin  temperatures. Spin temperatures measured as $<T_{s}>$ ($>3\sigma$) are represented by colored circles, where color indicates temperature (color scale, bottom left) and circle size represents the S/N of the measurement. The background gray scale map is as described for Figure \ref{fig:tau_map}.
\label{fig:Ts_map}}
\end{figure}

Some of the lowest $<T_{s}>$ values of $\sim{30}$ K are found in the outskirts of the SMC, as seen in Figure \ref{fig:Ts_map}, which suggests that even when there is less \hi\ there can still be cold \hi\ that dominates the weighted mean over the WNM. There are two measured values and three lower limits of $<T_{s}>$ that are significantly higher ($\gtrsim{200-300}$ K) than the main distribution of temperatures: the highest values of  $<T_{s}>$ are $\sim{800}$ K for source 004047-714559, but is only measured at $3\sigma$, and $\sim{450}$ K for source 011049-731427. The high average spin temperatures are primarily found along lines of sight with high column densities and little to no absorption suggesting that they indicate low CNM fractions. Generally, the spatial distribution of $<T_{s}>$ values shows no pattern and there are variations both on small and large scales. The lack of any strong trend of  $<T_{s}>$ with the column density (Figure \ref{fig:Ts_avg}) shows that the balance of the CNM and WNM component temperatures and fractions does not depend on the the total amount of \hi\ along the line-of-sight.

\subsection{Individual Cold \hi\ Cloud Temperatures}\label{sec:abs_comp}

By identifying individual absorption line features we can estimate the temperatures of the different velocity components of the absorbing \hi, which can be interpreted as individual cold \hi\ clouds that are surrounded by the warm inter-cloud \hi. We compare how the emission changes with respect to the absorption in order to disentangle the cloud emission from the surrounding warm-phase emission for a given absorption line. This method uses a linear fit to the emission-absorption plot as developed by \citet{dus96}, described by \citet{meb97}, and used on the previous SMC absorption line survey by \citet{dic00}. One of the benefits of fitting absorption features in the emission-absorption diagram is that we are able to estimate a cloud temperature directly, without making assumptions about the line profile shape or the number of blended components. This method assumes that emission is comprised of three components: the fraction, $q$, of warm gas that is in front of the cold cloud and unaffected by self-absorption; the emission from the cold cloud; and the remaining emission from the warm gas behind the cold cloud that is attenuated by it. In terms of the radiative transfer, we then have:
\begin{equation}
T_{B}(\text{v}) = qT_{ew}(\text{v}) + T_{ec}(\text{v}) + (1-q)T_{ew}(\text{v})e^{-\tau(\text{\text{v}})}
\end{equation}
where $T_{ew}(\text{v})$ is the emission brightness temperature of the warm gas and $T_{ec}$ is the emission brightness temperature of the cold cloud. Rearranging and putting this equation in terms of the spin temperature of the cold cloud ($T_{c}$) using $T_{ec}(v) = T_{c}(1-e^{\tau(\text{v})})$ gives
\begin{equation}
T_{B}(\text{v}) = (1-e^{\tau(\text{v})})\left[T_{c}-T_{ew}(\text{v})(1-q)\right] + T_{ew}(\text{v}).
\end{equation}
Then if we plot the absorption, $(1-e^{-\tau(\text{v})})$, as a function of the emission, $T_{B}(\text{v})$, the slope is
\begin{equation}
m(\text{v}) = \frac{dT_{B}(\text{v})}{d(1-e^{-\tau(\text{v})})} = T_{c}-T_{ew}(\text{v})(1-q) 
\end{equation}
and the y-intercept is $T_{ew}$, the emission brightness temperature at the line center. 
From this the cloud spin temperature can be estimated by
\begin{equation}
T_{c} = m+T_{ew}(1-q).
\end{equation}
Frequently the brightness temperature varies over the velocity range of the absorption, particularly in the SMC where the warm emitting \hi\ extends across a large range of velocities and there is typically not an isolated emission peak associated with the absorption feature. The slope and intercept can still be estimated by fitting the ridge line, which we do by averaging the pairs of points on either side of the peak of the absorption feature. 

We created synthetic spectra designed to mimic the observations to check that this method can recover an input cloud temperature. We produce the synthetic spectra by assuming there is one cloud of cold, absorbing gas with a single temperature ($T_{c}=30$ K) and a wider gaussian emission component. We set the FWHM of the cold gas absorption and emission to the observed narrow width of 2 km s$^{-1}$ and the FWHM $=20$ km s$^{-1}$ and peak brightness temperature $T_{B}=40$ K and the typical emission spectrum $RMS=1$ K, which resembles the simple, lower $S/N$ emission profile seen for source 003824-742212. We find we are able to recover the correct cloud temperature even for the low optical depth absorption features with moderate $S/N$ where we were able to measure the slope and $T_{c}$, specifically with $\tau_{max}\sim0.3$ and $\sigma_{\tau}\sim0.03$ (e.g., source 011005-722647). The method fails to recover the correct input temperature when the line FWHM falls below the resolution of the absorption spectrum (in this case FWHM $<0.6$ km s$^{-1}$). In these cases the method systematically underestimates the cloud temperature. 

\begin{figure*}[t!]
\plotone{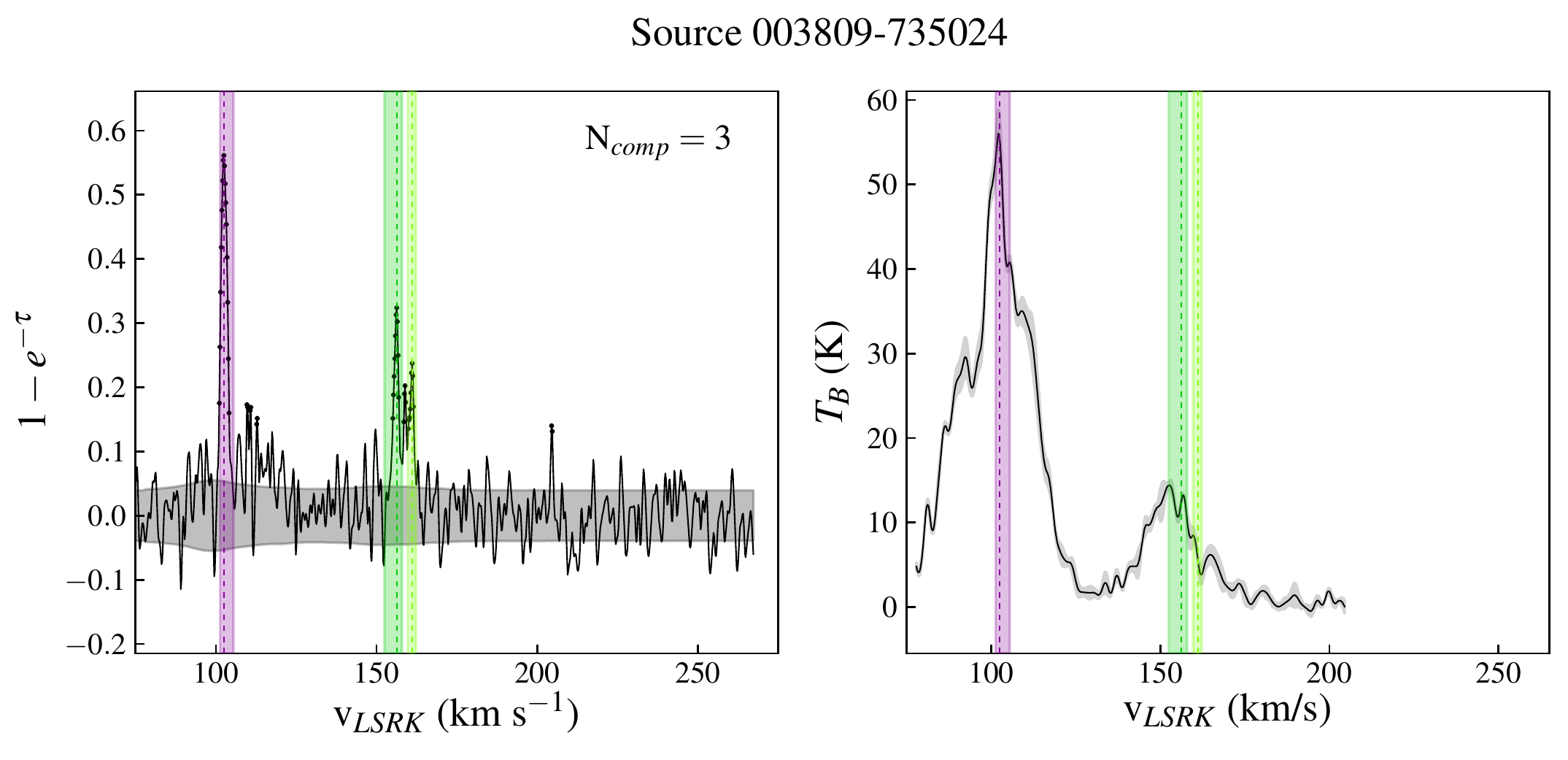}
\caption{Identified absorption features in absorption (\textit{left}) and emission (\textit{right}) spectra, source 003809-735024. Individual absorption features are identified by color, with the dashed line of the respective color used to indicate the peak of each feature. The number of components ($N_{comp}$) is indicated at the top right. The filled light gray area shows the $1\sigma$ uncertainty (see Section 3.1). The complete figure set (29 images) is available in the online journal.
\label{fig:abs_features}}
\end{figure*}

\begin{figure*}[t!]
\plotone{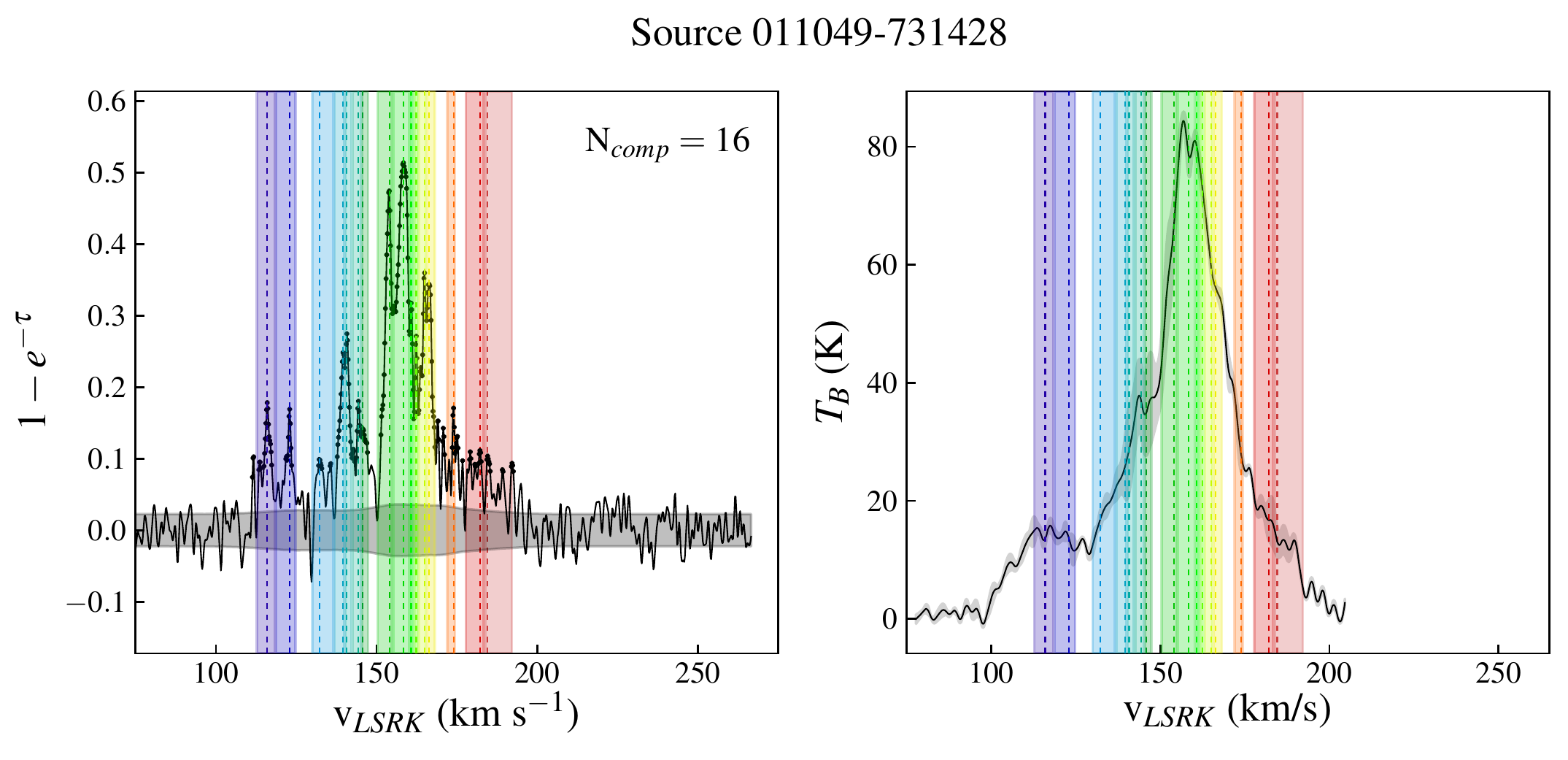}
\caption{Example of complex absorption spectrum, source 011049-731428. The many absorption features for this source are represented as described for Figure \ref{fig:abs_features}.
\label{fig:abs_features2}}
\end{figure*}

An alternative approach to using the emission-absorption diagram and fitting a ridge line is to decompose both the emission and absorption spectra into individual Gaussian components and compare the emission brightness temperature and opacity for a given component. Gaussian decomposition has been successful for simple, high signal-to-noise spectra, such as those found at high Galactic latitude (e.g., \citealt{hei03a}, \citealt{mur15}). However, \citet{mur17} showed that for noisy spectra with strong line blending, similar to many of the spectra in our sample, autonomous Gaussian decomposition of the spectra using Gausspy \citep{lin15} typically only recovered $\sim{50\%}$ of the actual line components. We attempted to decompose the spectra in an automated manner using GaussPy, but we were not able to reliably train the algorithm using synthetic training sets and the best accuracies reached were $\sim{60-70\%}$. More work will need to be done to try to remove noise from the spectra before fitting Gaussians and to better understand the systematic uncertainties. 

While \citet{hei03a} primarily used Gaussian decomposition to measure the CNM cloud temperatures, they found that the ridge line fitting method applied to a subset of line profiles produced comparable temperature estimates to those found with Gaussian decomposition. This shows that for spectra where both Gaussian decomposition and the emission-absorption diagram can be used, the two methods are not systematically different.

\begin{deluxetable*}{ccccccccccc}
\tablecaption{Absorption Feature Measurements \label{table:abs_features}}
\tabletypesize{\footnotesize}
\tablehead{
\colhead{Source Name} & \colhead{Component} & \colhead{Velocity Range} & \colhead{v$_{max}$} & \colhead{$(1-e^{-\tau})_{max}$} & \colhead{$m$}  & \colhead{T$_{ew}$} & \colhead{T$_{c}(q=0.5)$}  \\
\colhead{} & \colhead{} & \colhead{(km s$^{-1}$)} & \colhead{(km s$^{-1}$)} & \colhead{} & \colhead{(K)} & \colhead{(K)} & \colhead{(K)}
} 
\startdata
002906-735333 & 1 & $104.4-108.4$ & 106.4 & 0.7 & 20.13$\pm$1.50 & 20.75$\pm$0.87 & 30.51$\pm$1.62 \\
003809-735024 & 0 & $101.1-105.7$ & 102.5 & 0.6 & 25.61$\pm$1.22 & 41.16$\pm$0.56 & 46.19$\pm$1.28 \\
003824-742212 & 0 & $99.1-105.1$ & 103.3 & 0.1 & -1.33$\pm$0.49 & 15.74$\pm$0.05 & \nodata \\
003824-742212 & 6 & $122.5-125.5$ & 123.5 & 0.1 & 37.70$\pm$5.47 & 9.09$\pm$0.49 & 42.25$\pm$5.48 \\
003824-742212 & 10 & $134.1-151.5$ & 149.3 & 0.7 & 2.10$\pm$0.81 & 39.02$\pm$0.36 & 21.62$\pm$0.85 \\
003824-742212 & 12 & $152.9-155.1$ & 154.1 & 0.2 & -9.05$\pm$1.70 & 29.93$\pm$0.25 & \nodata \\
004956-723554 & 1 & $138.6-143.4$ & 140.4 & 1.0 & -1.71$\pm$0.28 & 47.17$\pm$0.22 & \nodata \\
005238-731245 & 7 & $141.6-144.6$ & 144.1 & 0.8 & -0.55$\pm$0.09 & 107.13$\pm$0.06 & \nodata \\
005238-731245 & 15 & $153.9-158.5$ & 157.6 & 1.0 & -1.55$\pm$0.66 & 77.01$\pm$0.63 & \nodata \\
005238-731245 & 16 & $158.5-160.5$ & 159.3 & 1.0 & 1.12$\pm$0.43 & 78.88$\pm$0.39 & 40.56$\pm$0.51 \\
005238-731245 & 19 & $163.1-170.9$ & 168.8 & 0.8 & 1.44$\pm$0.63 & 57.78$\pm$0.45 & 30.33$\pm$0.71 \\
005238-731245 & 20 & $170.9-174.5$ & 173.9 & 0.9 & 0.63$\pm$0.12 & 37.26$\pm$0.10 & 19.26$\pm$0.14 \\
005238-731245 & 22 & $175.4-177.8$ & 177.3 & 0.7 & -0.67$\pm$0.04 & 17.88$\pm$0.03 & \nodata \\
005337-723144 & 6 & $127.0-131.0$ & 130.0 & 1.0 & 2.43$\pm$0.23 & 36.53$\pm$0.21 & 20.70$\pm$0.28 \\
005732-741243 & 0 & $137.3-142.1$ & 139.7 & 0.5 & -3.27$\pm$0.37 & 3.18$\pm$0.14 & \nodata \\
010029-713826 & 0 & $145.0-148.6$ & 147.0 & 1.0 & 19.83$\pm$1.63 & 60.89$\pm$1.52 & 50.28$\pm$1.95 \\
011005-722647 & 10 & $163.4-166.4$ & 164.8 & 0.3 & 8.18$\pm$0.77 & 27.40$\pm$0.21 & 21.88$\pm$0.78 \\
011005-722647 & 11 & $166.4-170.4$ & 168.8 & 0.3 & 5.20$\pm$0.14 & 24.04$\pm$0.04 & 17.22$\pm$0.14 \\
011005-722647 & 21 & $190.4-193.8$ & 192.0 & 0.2 & -13.73$\pm$2.0 & 29.13$\pm$0.34 & \nodata \\
011035-722807 & 15 & $175.8-180.6$ & 178.0 & 0.8 & 5.91$\pm$0.21 & 21.52$\pm$0.14 & 16.67$\pm$0.23 \\
011049-731428 & 1 & $112.6-118.6$ & 116.0 & 0.2 & -19.37$\pm$5.1 & 16.82$\pm$0.72 & \nodata \\
011049-731428 & 8 & $142.4-145.4$ & 144.4 & 0.2 & 12.00$\pm$7.03 & 34.80$\pm$1.03 & 29.40$\pm$7.07 \\
011049-731428 & 11 & $150.2-155.0$ & 154.0 & 0.5 & -6.72$\pm$0.93 & 69.30$\pm$0.39 & \nodata \\
011049-731428 & 12 & $155.0-160.4$ & 158.4 & 0.5 & -21.01$\pm$4.3 & 90.13$\pm$1.97 & \nodata \\
011056-731404 & 16 & $167.6-172.8$ & 171.4 & 0.8 & 4.93$\pm$0.18 & 37.33$\pm$0.12 & 23.60$\pm$0.20 \\
011132-730209 & 13 & $167.6-182.6$ & 173.2 & 0.9 & -9.46$\pm$2.79 & 80.19$\pm$2.22 & \nodata \\
011432-732142 & 2 & $131.1-135.1$ & 132.9 & 0.4 & -0.56$\pm$0.50 & 21.40$\pm$0.18 & \nodata \\
011432-732142 & 11 & $154.7-161.3$ & 158.1 & 1.1 & 8.21$\pm$1.00 & 94.36$\pm$0.74 & 55.39$\pm$1.13 \\
011432-732142 & 16 & $168.9-171.9$ & 169.9 & 1.0 & -8.70$\pm$1.31 & 86.84$\pm$1.26 & \nodata \\
\enddata
\end{deluxetable*}

\subsubsection{Identifying Absorption Features}\label{sec:id_abs_features}

Relative to the current study, \citet{meb97} and \citet{dic00} had low velocity resolution and lower sensitivity \hi\ absorption spectra, that generally only showed one or two identifiable absorption features, for each of which they were able to identify the separate ridge lines using the emission-absorption plots. The \hi\ absorption spectra in our sample show many more features and necessitate the identification of individual absorption features before the production of emission-absorption plots. To define our absorption features, we divide the spectra into sections based on the presence of absorption peaks. 

We find local maxima within these sections, from the $1-e^{-\tau}$ plots, so that the absorption is positive, and amongst the channels with $>3\sigma$ detection. We choose only the maxima that are different by $>3\sigma$ from either of the local minima found to each side of them. An absorption feature is defined as all of the data points between the two local minima surrounding a local maximum. Figure \ref{fig:abs_features} illustrates this method, showing identified absorption features in corresponding absorption and emission spectra. Figure \ref{fig:abs_features2} shows an example of a more complex spectrum. In this way, we identify at least one absorption feature in 29 sources out the 35 total sources with detected absorption based on the equivalent width measurement.

\subsubsection{Temperatures of Absorption Features}\label{subsec:abs_temps}s

\begin{figure*}[t!]
\plotone{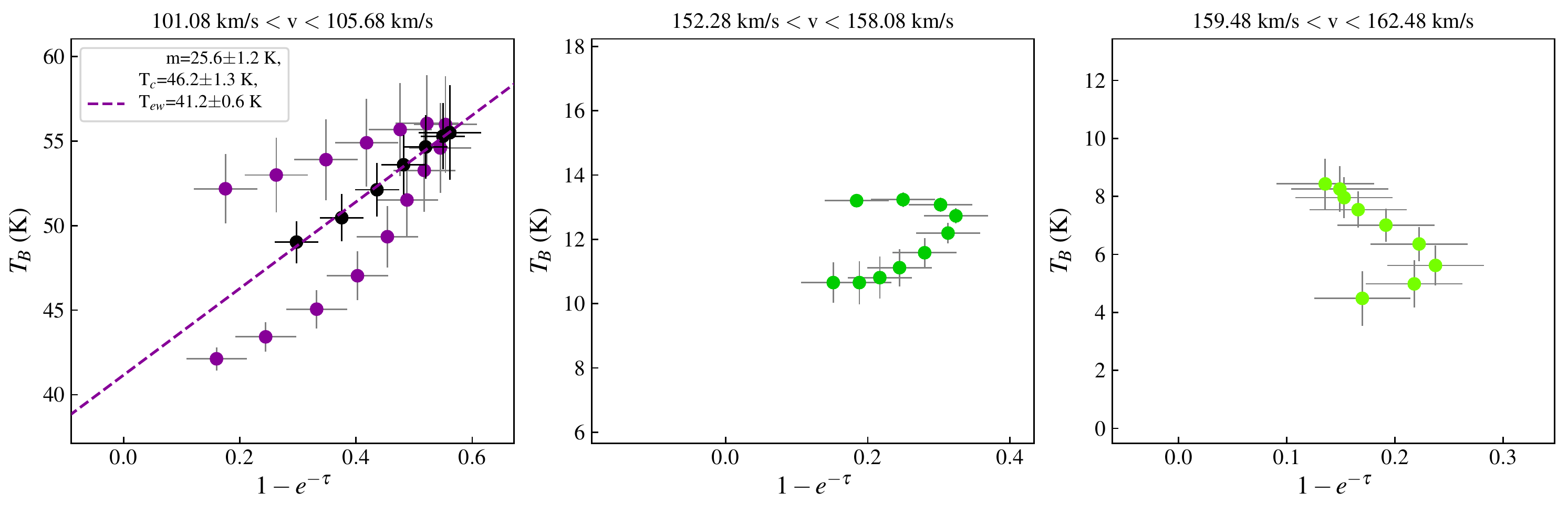}
\caption{Emission vs. absorption plots for identified absorption features, source 003809-735024.  Filled colored circles represent the absorption features shown in the same colors in Figure \ref{fig:abs_features};  velocity ranges are indicated above the plots. The gray error bars show the $1\sigma$ uncertainty. A single absorption line will trace out an arc where the slope and the intercept can be used to estimate the temperature of the cold gas (see Section \ref{subsec:abs_temps} for a detailed description). If an absorption feature has more than 3 points detected at $>3\sigma$ on either side of the absorption peak then we step out in equal increments in velocity space and average the points to be able to fit a ridge line to the arc; these average points are shown in black with the error. The dashed colored line shows the fit to the averaged points using least squares regression. The complete figure set (26 figures) is available in the online journal.
\label{fig:em_abs_features}}
\end{figure*}

\begin{figure*}[t!]
\plotone{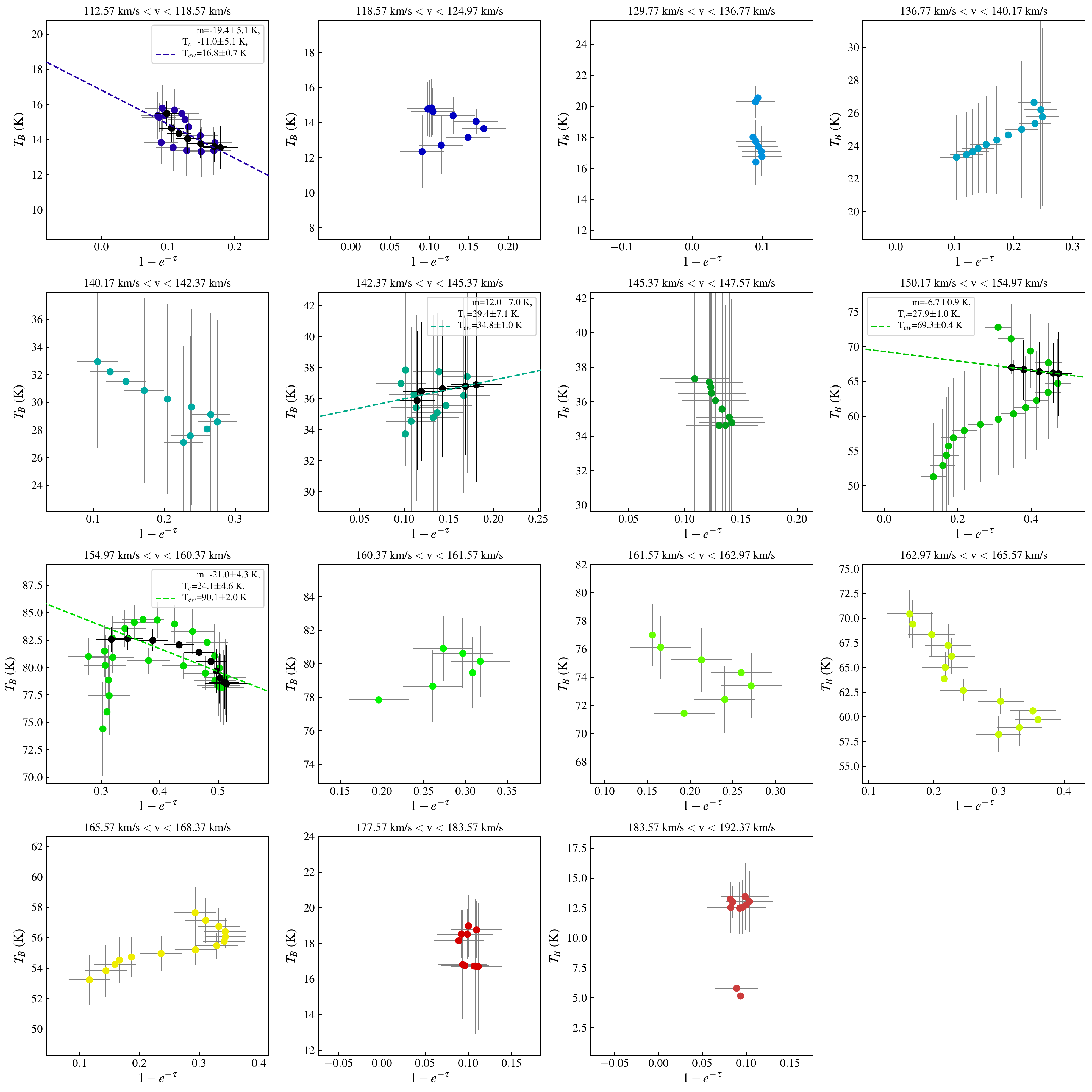}
\caption{Emission vs. absorption plots for identified absorption features, source 011049-731428.  Filled colored circles represent the absorption features shown in the same colors in Figure \ref{fig:abs_features2}; the data are presented as described for Figure \ref{fig:em_abs_features}.
\label{fig:em_abs_features2}}
\end{figure*}

\begin{figure*}[t!]
\plotone{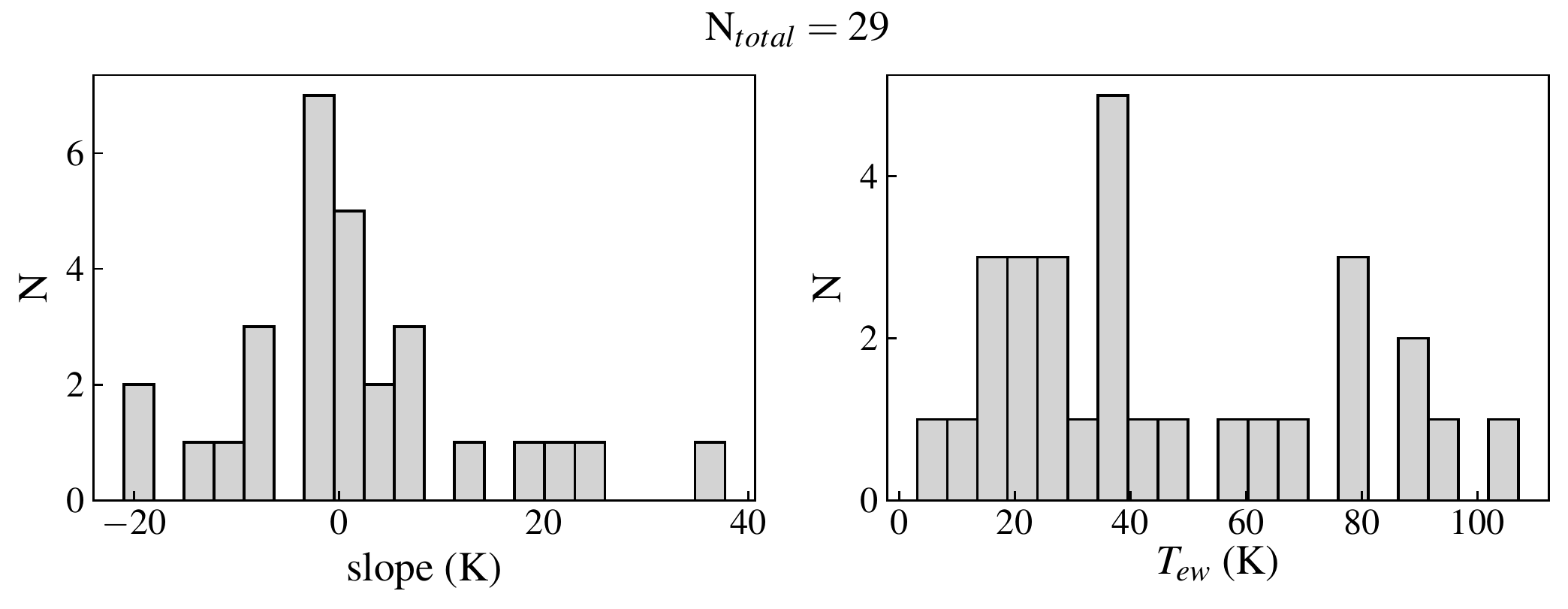}
\caption{Histograms of fitting results from emission-absorption diagrams for absorption features with more than four points along their ridge lines. {\it Left:} values of the slope ($m= T_{c}-T_{ew}(\text{v})(1-q)$) where $m>1\sigma$. Negative values of $m$ indicate self-absorption. {\it Right:} the intercept corresponding to the emission brightness temperature associated with the absorption feature ($T_{ew}$).  
\label{fig:em_abs_hist}}
\end{figure*}

Figure \ref{fig:em_abs_features} shows the emission-absorption plots for the absorption features identified in the spectrum of source 003809-735024 with data points corresponding to those in the left panel of Figure \ref{fig:abs_features}. Figure \ref{fig:em_abs_features2} shows the emission-absorption plots associated with the absorption features from the more complex spectrum seen in Figure \ref{fig:abs_features2}. When there are at least 8 detected data points (at $\geq{3\sigma}$) that are monotonically increasing towards and then monotonically decreasing away from the absorption peak, we create a ridge line and use  a linear least-squares regression to fit a slope and intercept to it. Table \ref{table:abs_features} lists the properties of all the source absorption features where there were enough detected data points to create a linear fit of the ridge line (29 out of 159 total identified absorption features).  Because we are only fitting the peaks in the absorption, this method will be biased towards the most optically thick, and probably coldest, \hi\ velocity components or clouds. 

Figure \ref{fig:em_abs_hist} shows the distribution of fitted slopes and the warm-phase brightness temperature. In the emission-absorption plots, strong self-absorption in the emission spectrum will manifest as negative slopes (a decrease in the emission $T_{B}$ associated with the absorption line). Figure \ref{fig:em_abs_features2} shows examples of fitted negative slopes for the absorption features at v $\sim{116}$ km s$^{-1}$ (top row, far left), v $\sim{154}$ km s$^{-1}$ (second row from top, far right), and v $\sim{158}$ km s$^{-1}$ (third row from top, far left). Figure \ref{fig:em_abs_hist} demonstrates a distribution about zero in the slope values; $48\%$ of the absorption features have negative slopes, indicating that cold gas clouds are located in front of the primarily warm gas producing the emission.

\begin{figure}[t!]
\epsscale{1.1}
\plotone{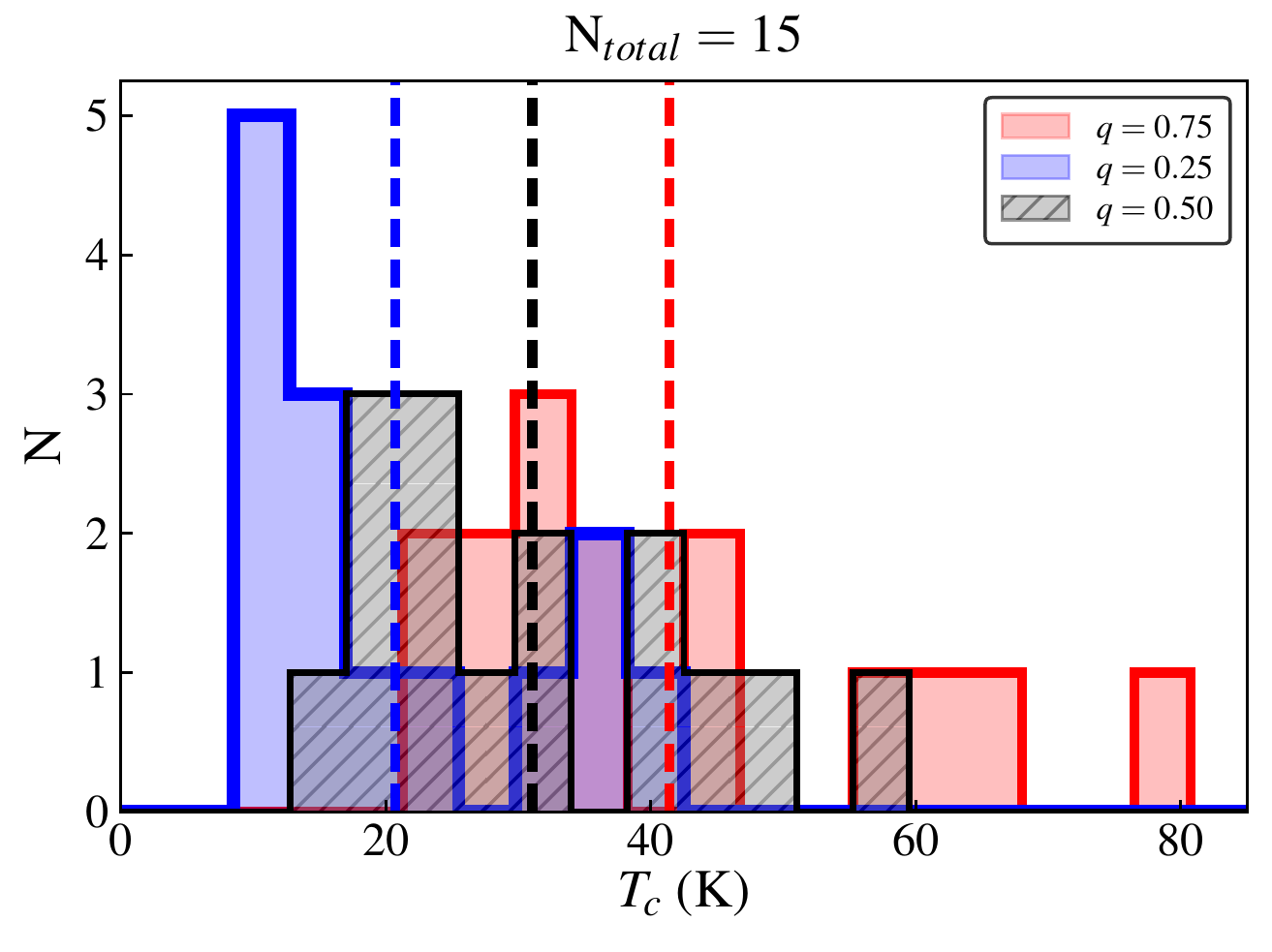}
\caption{Cold gas temperature estimates. Histograms of the cold gas temperatures estimated by fitting  ridge lines to the emission-absorption diagrams, using  different assumptions of the distribution of the cold \hi\ gas with respect to the warm: $q$ indicates the fraction of the cold absorbing gas in front of the warm medium. The vertical dashed lines show the mean $T_{c}$ values of 21 K, 31 K, and 41 K for $q=0.25$, $q=0.5$, and $q=0.75$, respectively.
\label{fig:Tc_hist}}
\end{figure}

In converting the fitted slope and intercept ($T_{ew}$) to the spin temperature of the cold cloud ($T_{c}$), we must make an assumption about the fraction of warm-phase gas that is in front of the absorbing \hi\ cloud. In Table \ref{table:abs_features} we give the $T_{c}$ values assuming $50\%$ of the warm gas is in front of the cold ($q=0.5$), for simplicity and to be directly comparable to the previous results for the SMC \hi\ absorption survey by \citet{dic00}. Given the prevalence of self-absorption, the fraction of warm gas in front of the cold gas, $q$, must in general be low. Figure \ref{fig:Tc_hist} shows the distribution of $T_{c}$ for individual absorption features for a range of fractions of warm gas in front of the absorbing gas. The results for different fractions ($q$) change the distributions, but the effect is not dramatic. The average $T_{c}$ values range from 21-41 K for $q=0.25-0.75$. Given that the typical uncertainties on the $T_{c}$ estimates are $\sim{1-5}$ K, the average for $q=0.5$ of $T_{c}\sim{30}$ K is a reasonable value for the overall average temperature of cold \hi\ absorption features.

Figure \ref{fig:Tc_Ncor} and \ref{fig:Tc_peak_tau} show flat distributions of the cloud temperatures with column density and the peak optical depth of the absorption feature. This means that there is no strong trend of $T_{c}$ in relation to the optical depth of the cloud or the total amount of \hi\ gas along the line-of-sight in which the cloud is located. Also, this indicates that the average $T_{c}$ for our subsample of features with temperature measurements is unlikely strongly biased because they cover a wide range in optical depth and column density. There is, however, what appears to be a floor in the cold cloud temperatures at $\sim{15\pm5}$ K, with the range due to variance from different $q$ values. Figure \ref{fig:Tc_peak_tau} also demonstrates that our $T_{c}$ measurements are not biased with respect to the optical depth of the absorption feature.

When considering the location of the various cloud temperatures, Figure \ref{fig:Tc_maps} demonstrates the lack of any strong trend of $T_{c}$ with the location within the galaxy both in terms of the spatial location and velocity of the feature. Similarly, the locations of features with negative slope indicating self-absorption (black points) do not show any strong trend with velocity, suggesting that cold gas that is able to produce self-absorption is randomly distributed throughout the emitting, warm gas. 

\subsection{Fraction of Cold \hi\ Gas}\label{subsec:fc}

Given the model of a two-phase neutral ISM composed of a cold and warm component, we can combine our measurements of $<T_{s}>$ and $T_{c}$ to estimate the average fraction of cold neutral gas ($f_{c}$) in the SMC. Rearranging Equation \ref{eqn:fc}, the cold gas fraction can be estimated by
\begin{equation}
f_{c}\simeq{\frac{T_{c}}{<T_{s}>}}.
\end{equation}
Using the representative cloud temperature of $T_{c}=30$ K as the characteristic temperature of the cold phase of the neutral gas and the column density weighted average spin temperature of $<T_{s}>=150$ K, we find $f_{c}\sim{20\%}$ in the SMC based on the lines of sight observed in this sample.  

\section{Discussion}\label{sec:discussion}

\subsection{Emergence of Narrow and Shallow and Wide Absorption Features}

Due to the increased velocity resolution of the current survey relative to earlier ones, we now see that many of the spectra show very narrow absorption line profiles that were previously unresolved. All of the high optical depth line features are narrow, with FWHM $\sim{2}$ km s$^{-1}$ (e.g. Figure \ref{fig:src_abs}), and the spectra generally appear multi-peaked and jagged (e.g., Figure \ref{fig:em_abs_features2}), which may arise from the superposition of multiple narrow absorption line components associated with kinematically distinct \hi\ gas clouds along the line of sight. The narrowness, and often high optical depth, of components would be expected for cold gas temperatures. 

A number of spectra also show wide, low optical depth absorption components. The range of velocities covered by the wide components coincides with that of the \hi\ emission profile. The spectrum shown in Figure \ref{fig:src_abs} shows an example of a wide, low optical depth absorption component with a width $\sim{20}$ km s$^{-1}$ and peak optical depth of $\tau\sim{0.2}$ covering velocities from $\sim{90-125}$ km s$^{-1}$, which covers most of the range of the lower velocity emission features. One possibility is that these are tracing warmer neutral gas components with low to moderate optical depth. However, for source 003824-742212 (shown in Figure \ref{fig:src_abs}), if we attributed all of the absorption ($EW=5.7$ km s$^{-1}$) to a warm phase and estimated the temperature based on the associated column density from the emission, then we obtain $T\sim{200}$ K. The moderate levels of optical depth preclude the possibility of true WNM ($T>5000K$) or even intermediate phase gas ($T>500$ K). These wide features are most likely a superposition of many narrow line components, similar to the narrow high optical depth absorption features we see in the spectra. Because of the low $S/N$, it is difficult to determine the exact width (and therefore temperature) of the individual low optical depth absorbing components. 


\begin{deluxetable}{ccc}
\tablecaption{Comparison of this survey to \citet{dic00} \label{table:compare_dic00}}
\tablehead{
\colhead{} & \colhead{\citet{dic00}} & \colhead{This Work} 
} 
\startdata
$\sigma_{\tau}$ & $0.05-0.2$\tablenotemark{a} & $0.01-1.0$\tablenotemark{b} \\
$\Delta\text{v}$ & $0.8$ km s$^{-1}$ & $0.2$ km s$^{-1}$ \\
$N_{src}$ & 28 & 55 \\
$N_{abs}$ & 13 & 37 \\
\enddata
\tablenotetext{a}{in spectra with $0.845$ km s$^{-1}$ wide channels}
\tablenotetext{b}{in spectra smoothed to $0.6$ km s$^{-1}$ with $0.2$ km s$^{-1}$ wide channels}
\end{deluxetable}

\subsection{Distribution of Cold \hi\ Gas}\label{subsec:distribution_cold}

One of the most striking results from this survey is the extent to which we see \hi\ absorption throughout the  galaxy, including outside the main body of the SMC. Figures \ref{fig:tau_map} and \ref{fig:Ts_map} show that we see absorbing, optically thick and primarily cold \hi\ throughout the SMC and in both high and low column density gas. Figure \ref{fig:Tc_maps} shows that there are self-absorption sources (black points) distributed across the whole range of velocities where there is \hi\ emission.  There is no obvious spatial grouping of the \hi\ gas of a given optical depth or average spin temperature, although the spectra with lower peak $\tau$ tend to be located outside the main body of the galaxy. Similarly, the distribution of cloud temperatures for individual absorption features shows no clear trend with spatial location or velocity. 

Because self-absorption requires there to be optically thick gas in front of the warm, emitting gas, the distribution of self-absorption has the potential to help disentangle the three-dimensional structure of the gas. Despite the fact the \hi\ emission line profiles are extended in velocity typically covering $\sim{100}$ km s$^{-1}$, we see absorption features across the whole velocity range. Physically, the optically thick, mostly cold \hi\ gas clouds appear to be distributed throughout the warmer, more extended \hi\ responsible for the emission. 

\begin{figure}[t!]
\epsscale{1.1}
\plotone{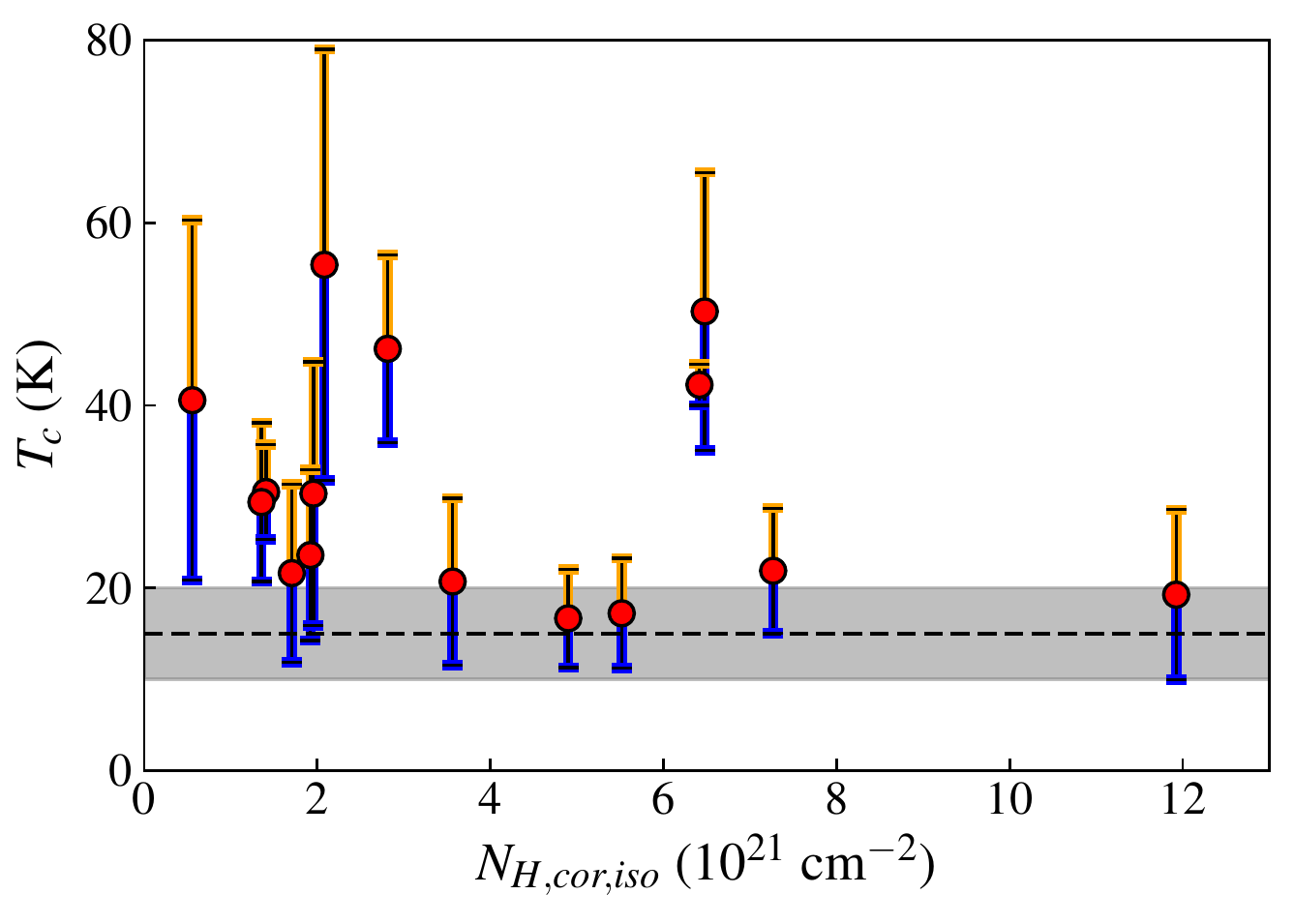}
\caption{Cold \hi\ gas temperature ($T_{c}$) vs. total \hi\ column density. The red points show the $T_{c}$ estimate for $q=0.5$ and the error bars show the range of possible $T_{c}$ from assuming $q=0.25$ (lower limit in blue) and $q=0.75$ (upper limit in orange). The dashed line shows $T_{c}=15$ K with the gray shaded area showing $\pm5$ K that accounts for the range of values based on assumed $q$, which appears to be a floor in the cold gas temperatures.
\label{fig:Tc_Ncor}}
\end{figure}

\begin{figure}[t!]
\epsscale{1.1}
\plotone{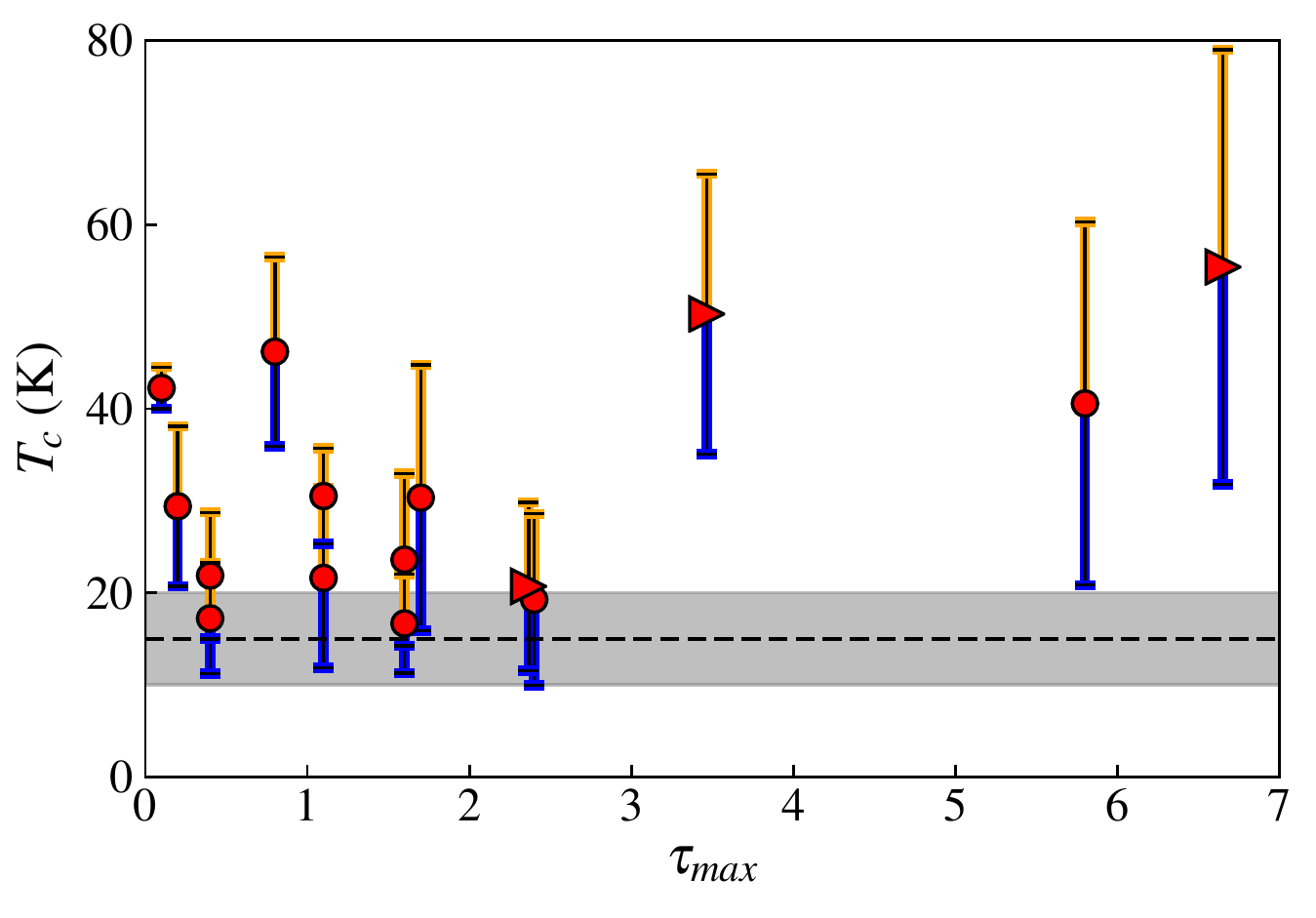}
\caption{Cold \hi\ gas temperature ($T_{c}$) vs. maximum optical depth ($\tau_{max}$).  Filled red circles show the cold \hi\ gas temperature ($T_{c}$) estimate for individual absorption features assuming $q=0.5$. Error bars show the range of possible $T_{c}$ for $q=0.25$ (lower limit in blue) and $q=0.75$ (upper limit in orange). The right-facing triangles indicate where the measured $(1-e^{-\tau})_{max}>1.0$ due to noise; in these cases, peak $\tau$ for the entire spectrum is used as a lower limit on the $\tau_{max}$ for each absorption feature. The dashed line shows $T_{c}=15$ K, and the gray shaded region shows $15\pm5$ K, which appears to be a floor in the cloud temperature estimates.
\label{fig:Tc_peak_tau}}
\end{figure}

\begin{figure*}[t!]
\epsscale{1.1}
\plotone{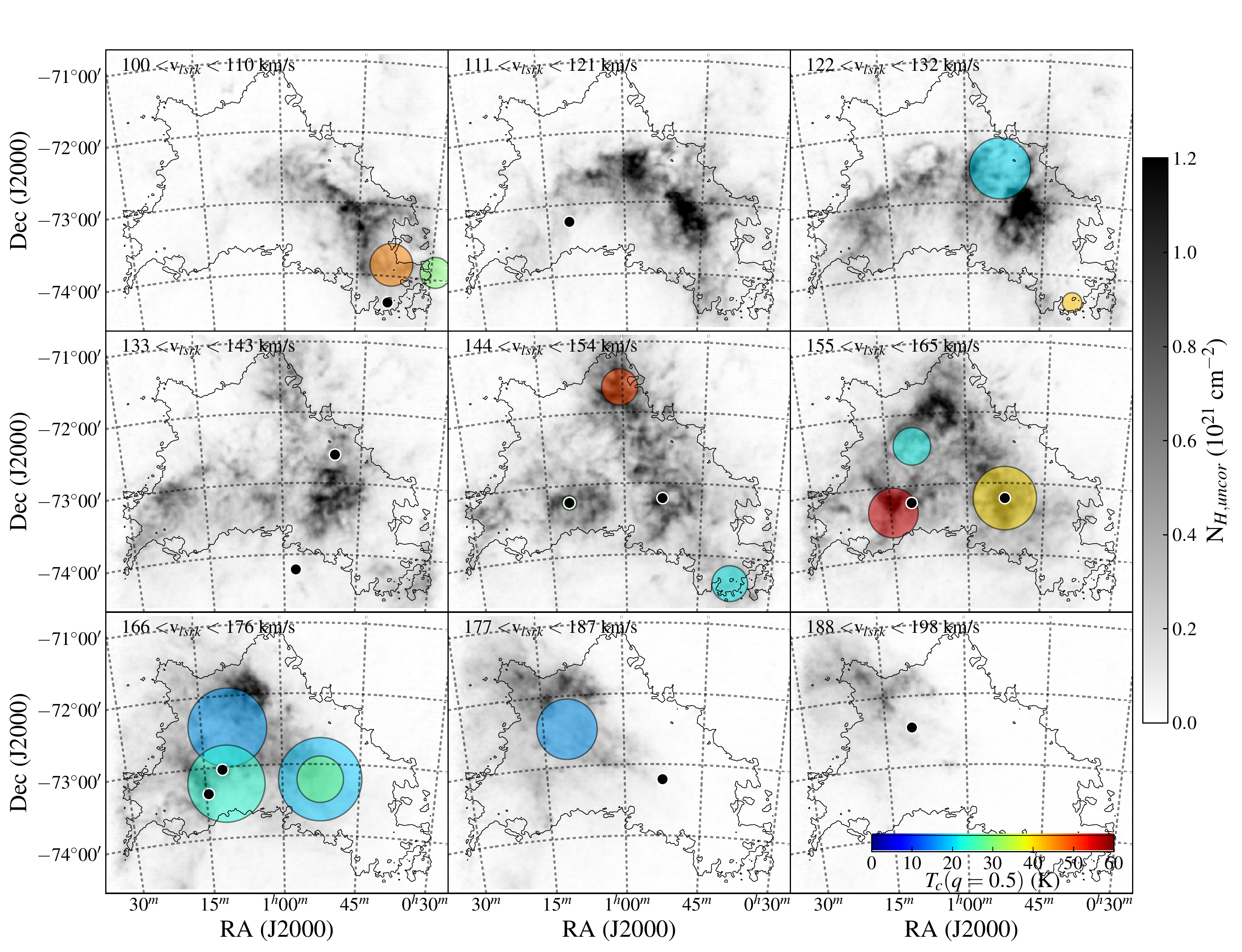}
\caption{Spatial and velocity distribution of cold \hi\ gas components. Cold gas components are represented by colored circles, where color indicates temperature (color scale, bottom right) and size represents the relative $S/N$ of the measurements. The \hi\ emission has been split into integrated intensity maps covering 10 km s$^{-1}$ with each absorption component shown on the corresponding map for the velocity of the absorption peak (v$_{max}$). Absorption components with negative slopes indicating self-absorption are shown with black points. The detected absorption components tend to be co-located with the emission. The background gray scale image shows the column density of \hi\ (assuming optically thin emission), and the main body of the galaxy is indicated by the black contour line, representing $N_{\text{H,uncor}}=2\times10^{21}$ cm$^{-2}$. 
\label{fig:Tc_maps}}
\end{figure*}

While Figures \ref{fig:tau_peak} and \ref{fig:Ts_avg} show that there is no strong correlation between the peak optical depth or the average spin temperature and the total \hi\ column density, the highest column densities also have some of the highest values of $\tau_{peak}$. Again, we see no relationship between cold cloud temperature and total amount of \hi\ along the line or sight or the maximum optical depth of the absorption feature in Figures \ref{fig:Tc_Ncor} and \ref{fig:Tc_peak_tau}. 

\subsection{Comparison to Previous Work}\label{subsec:previous_work}

The first large survey of \hi\ absorption in the SMC, which also used the ATCA, was presented in \citet{dic00}. Table \ref{table:compare_dic00} summarizes the improvements from the previous survey in terms of the optical depth sensitivity ($\sigma_{\tau}$), velocity channel size ($\Delta \text{v}_{ch}$), number of detected continuum sources ($N_{src}$), and the number of sources with detected absorption ($N_{abs}$). With the new survey, we achieve up to $5\times$ better optical depth sensitivity and $4\times$ better velocity resolution. We detect $\sim{2}\times$ as many continuum sources and $\sim{3}\times$ as many sources with detected \hi\ absorption.

We now detect absorption in 7 out of the 10 sources we re-observed from the previous survey that then had no detectable absorption, which we attribute to the enhanced optical depth sensitivity and velocity resolution of our new survey. For 15 out of the 21 sources that overlap with the \citet{dic00} survey, we measure $<T_{s}>$ that is inconsistent with the previous estimate and all our measured $<T_{s}>$ values are systematically lower. Appendix \ref{appendix:previous_survey} describes in detail the differences in how the absorption spectra were extracted in \citet{dic00}, which explains this systemic difference. In particular, the lower $<T_{s}>$ measurements are found for sources with narrow, deep absorption lines and/or wide, low optical depth (shallow) absorption line features, which the improved instrumentation and calibration would allow us to detect. Both higher measured $\tau$ and detection of shallow features increase the measured equivalent width of the absorption line spectra, which decreases the estimate of $<T_{s}>$. The full comparison of all the $<T_{s}>$ and optical depth sensitivities for sources that were re-observed can be found in Appendix \ref{appendix:previous_survey} in Table \ref{table:compare_Ts}.

Despite significant differences in the measured $<T_{s}>$ values for the same individual absorption line components that are identified in both samples, we find similar cloud temperatures ($T_{c}$), using the same method of fitting the ridge line in the emission-absorption diagram. For example, the temperature for the primary absorption feature in source 005337-723144 is $T_{c}=20.6\pm1.7$ K compared to $T_{c}=23\pm9$ K measured by \citet{dic00}. Detailed comparison for all sources that were observed in both surveys is not straightforward, given the increased complexity of the new spectra and the consequent difficulty in comparing the same absorption features.  In spite of this, \citet{dic00} did find a similar typical cloud temperature of $T_{c}\sim{20-30}$ K and range of temperatures from $10-70$ K. The similarity is not surprising, since  we are only able to produce and fit a ridge line to absorption features with enough points on either side of the absorption peak; this generally selects the strongest absorption features, which would also have been the dominant features in the previous survey spectra. Our measurement of similar temperatures is supported by the improved sensitivity of our data and the greater number and precision of our measurements.

In general, even less sensitive surveys can accurately estimate the cloud spin temperatures for strong absorption features that are isolated using the method of fitting the ridge line in the emission-absorption plot. The enhanced sensitivity we achieve here can lead to a larger number of individual cloud temperature measurements, which can change the estimated average cold cloud temperature, as well as improving the accuracy of the average spin temperature measurements. Combining these two factors means that while our estimate of the average temperature of individual clouds may not change, our measurement of the average cold gas fraction might. In the case of the SMC, \citet{dic00} estimated $f_{c}<15\%$, whereas the increased sensitivity of this survey raises the fraction to $f_{c}\sim{20\%}$ based on the new sample.

\subsection{Comparison to the Milky Way}

While we reserve detailed comparison of the SMC results for a forthcoming paper that includes the results from of the LMC, it is worthwhile to compare to the existing comparable studies of the Milky Way. We see that the peak optical depths are generally higher than those found in the Perseus molecular cloud region \citep{sta14} and in random lines of sight throughout the Milky Way, although certain lines of sight can reach as high as $\tau_{max}\sim{8}$ \citep{hei03a}. We find a similar fraction of optically thick lines of sight as was found for the Riegel-Crutcher cloud \citep{den18}.


Taking our average estimate of the cold gas fraction in the SMC of $\sim{0.20}$, we see that it is similar to median CNM fraction of 0.33 in Perseus \citep{sta14} and $\sim{0.15-0.2}$ throughout the Milky Way \citep{hei03b,mur15,kol19}. Despite the lower metallicity environment in the SMC, there is a similar fraction of CNM to WNM as found in the Milky Way, and there are very optically thick \hi\ components that are similar to those found around Milky Way molecular clouds, but the temperatures of individual \hi\ absorption components appear to be somewhat colder than those found at higher metallicity in the Milky Way. 

When we compare individual absorption component spin temperatures, the average for the SMC of $T_{s}\sim30$ K based on the subset of sources where the cloud temperatures can be measured is lower than that found for the Riegel-Crutcher cloud (48 K), for Perseus (42 K), and throughout the Milky Way (44 K) if you only consider comparable higher optical depth lines ($\tau>0.2$). If we compare the CNM temperature estimates for absorption line components from \citet{hei03a} where they estimate the temperatures using a similar slope fitting method, the temperatures span a similar range, but the median temperature of 43 K is still higher than our finding from our sample of sources in the SMC. 



\subsection{Very Cold \hi\ Cloud Temperatures in the SMC}

The average cold \hi\ cloud temperature that we measure here is $T_{c}\sim30$ K, which is robust to the uncertainty from the assumed location of the warm gas with respect to the cold. This temperature is also noticeably lower than that found for cold cloud temperatures measured in the Milky Way. The coldest temperatures ($T_{c}\lesssim20$ K) are reached even at low cloud optical depths (Figure \ref{fig:Tc_peak_tau}), which suggests that it is not local shielding that determines the temperature of the clouds. These lowest cloud temperatures have a tendency to be located along lines of sight with high \hi\ column densities ($N_{\text{H,cor,iso}}\gtrsim5\times10^{21}$ cm$^{-2}$), which may indicate that the aggregate shielding along the line of sight does affect the ability of the \hi\ to cool to low temperatures in the SMC. 

\section{Conclusions}\label{sec:summary}

We present a new, large SMC ATCA survey of \hi\ absorption spectroscopy against background radio continuum sources. Smoothed versions of the spectra that maintain a relatively high velocity resolution of 0.6 km s$^{-1}$ are used for our analysis. We find evidence of absorption in 37 out of a total of 55 total detected continuum sources based on the measured equivalent widths of the absorption spectra. We further split the spectra into distinct absorption features to find the properties of different velocity components of the absorbing \hi\ gas. Absorption features are identified by finding $3\sigma$ differences between local minima and maxima in the smoothed spectra and cold cloud temperatures are estimated from the data using the method of fitting slopes to the emission-absorption diagrams. 

We draw two main conclusions from our sample from the initial analysis of the new survey of \hi\ absorption in the SMC:
\begin{enumerate}
\item{The cold \hi\ gas we observe is distributed throughout the SMC both spatially and in velocity. There are no strong trends of temperature with column density or region of the galaxy for our sample. The \hi\ absorption velocity components with self-absorption are also distributed throughout the galaxy and appear across most of the range of velocities with emission.}
\item{From measurements of individual absorption components for a subset of our sample of sources, we find the average cold \hi\ cloud temperature is $T_{c}\sim{30}$ K in the SMC, which is lower than the $\sim{40}$ K found in the Milky Way. Using the column density weighted average spin temperatures ($<T_{s}>$) of $150$ K for our entire sample gives an estimate of the cold gas fraction in the SMC of $\sim{20\%}$, which is similar to the average value for the Milky Way but larger than previously estimated for the SMC.}
\end{enumerate}

\acknowledgements

The Australia Telescope Compact Array is part of the Australia Telescope National Facility which is funded by the Australian Government for operation as a National Facility managed by CSIRO. K.J. and N.Mc.-G. acknowledge funding from the Australian Research Council via grant FT150100024. D. L. wishes to acknowledge support from National Key R\&D Pro-gram of China No. 2017YFA0402600, the CAS International Partnership Program No. 114A11KYSB20160008. JRD is the recipient of an Australian Research Council (ARC) DECRA Fellowship (project number DE170101086). Parts of this research were supported by the Australian Research Council Centre of Excellence for All Sky Astrophysics in 3 Dimensions (ASTRO 3D), through project number CE170100013. \textbf{Finally, we wish to thank the anonymous referee whose comments improved the quality of this manuscript.}

\facilities{ATCA}.

\software{NumPy \citep{oli06}, SciPy \citep{jon01}, Astropy \citep{astropy:2018,astropy:2013}, Matplotlib \citep{hun07}, NADA \citep{lee17}}

\appendix

\section{Detailed Comparison to Previous ATCA \hi\ Absorption Survey of the SMC}\label{appendix:previous_survey}

Table \ref{table:compare_Ts} presents a comparison of the sensitivities and average spin temperature ($<T_{s}>$) for all sources that overlap with the previous survey by \citet{dic00}. For each source where the $<T_{s}>$ measurements are discrepant, the Notes column indicates the reason for the differing values: the new spectra showed either or both shallow (low-$\tau$) and wide absorption features or narrow and deep (high-$\tau$) absorption features that were not fully resolved previously.

To explain the discrepancy between the optical depths measured in the earlier study (data taken 1995 July) we have recovered the raw data from the ATCA archive, recalibrated and reprocessed the spectra in the direction of one of the strongest background sources, J003824-742212. In this case, and for several of the other sources on Table \ref{table:compare_Ts}, the 1995 data were biased toward lower values of optical depth, hence lower equivalent widths and higher spin temperatures compared with those measured in the same directions in this survey.  The cause of the bias appears to be in the analysis method used in the earlier survey.

The 1995 data were analyzed on the {\it{u,v}} plane, since each source had typically one hour or less integration time, so there was hardly enough data to make a map with reasonable dynamic range. The 15 baselines were analyzed separately, with spectra for each baseline computed by vector averaging the calibrated {\it u,v} data.  The resulting average spectra were then arithmetically averaged with weighting by the continuum flux measured on each baseline. Baselines shorter than 3 K$\lambda$ = 600 m were given zero weight if HI emission lines were evident, as they often were. The different baselines have different continuum levels because the background sources are typically partially resolved, particularly on baselines to antenna 6, with lengths 3 to 6 km. A spectral baseline was fitted to channels well away from the SMC velocity, and the spectra were divided by this continuum estimate to get spectra of $e^{-\tau}$, as in Figure \ref{fig:src_abs} (left side, Section \ref{sec:results} above).

The bias enters in the arithmetic averaging of the spectra from each baseline.  Curiously, although each of these 15 spectra is a noisy but unbiased estimate of the absorption, averaging them together creates a bias due to the presence of emission structure scattered around the primary beam area of the telescope ($32\arcmin$ beam width).  These low level emission fluctuations add incoherently in the vector averages used to construct the spectra for each baseline, but averaging the baselines is effectively a scalar average of different samples of {\it u,v} data.  The emission fluctuations always contribute positively to the scalar average, and thus counteract the effect of absorption.

\begin{figure}[ht]
\hspace{1.5in}\includegraphics[width=3.5in]{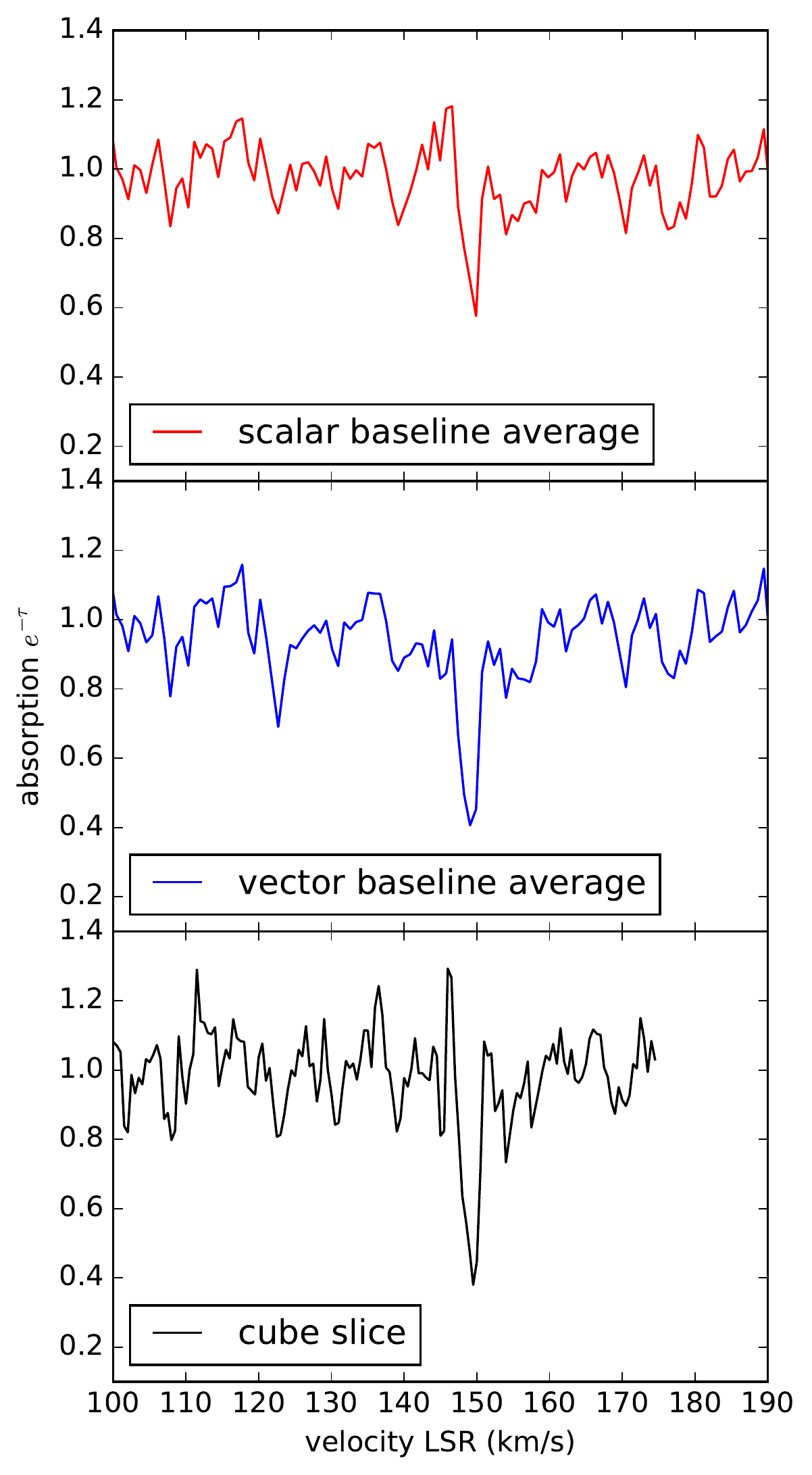}
\caption{Spectra toward background source J003824-742212 made with data presented in \citet{dic00}.  The top spectrum is an weighted average of the spectra measured on each interferometer baseline, computed using MIRIAD task UVSPEC.  It is biased by emission fluctuations away from the phase centre that do not fully cancel out in the scalar average.  The middle panel shows a weighted vector average of the spectra from each baseline (real part).  The bottom panel shows a spectrum sliced through a cube made by Fourier inversion of the {\it u,v} data.  The middle and bottom spectra are consistent with the spectrum in Figure \ref{fig:src_abs} measured in this survey.  Note that the noise in the two upper spectra is very similar at velocities away from the range of SMC \hi\ emission ($\sim{140-160}$ km s$^{-1}$ in this direction).
\label{fig:appendix_003824} }
\end{figure}

Mapping the {\it u,v} data, as done in this project, separates the fragments of emission that are scattered around the primary beam, giving a cleaner estimate of the spectrum toward the  background source.  To check this we compute spectral line cubes from the 1995 data, and  make spectra by slicing the cubes through the peak continuum pixel.  For J003824-742212 the result is shown on the bottom panel of Figure \ref{fig:appendix_003824}. This spectrum is in good agreement with the spectrum measured in this survey. The lowest point in the 1995 spectrum has $e^{-\tau}= 0.39\pm0.15$. J003824-742212 shows peak optical depth in this survey of  1.1, hence $e^{-\tau}$ = 0.33.

The top spectrum on figure \ref{fig:appendix_003824} is the result of weighted arithmetic averaging of the spectra for each baseline.  This absorption line is not as deep as in the bottom profile, $e^{-\tau} = 0.6 \pm 0.15$ instead of 0.4, giving $\tau$ = 0.51.  This is consistent with figure 2 in \citet{dic00}, that was computed the same way, with somewhat different weighting and smoothing.

It is interesting that although the individual baseline (vector-averaged) spectra that go into the average are not biased by the emission, their average is.  This was not expected in the analysis of the 1995 data.  Scalar averaging has insidious effects on spectra in the {\it u,v} plane.  The approach taken in this paper, doing the Fourier inversion to the sky or {\it l,m} plane and then slicing the resulting cube, is a safer method, but more computationally intensive.

\begin{deluxetable*}{cccccccc}
\tablecaption{Detailed comparison to \citet{dic00} \label{table:compare_Ts}}
\tabletypesize{\footnotesize}
\tablehead{
\colhead{Field} & \colhead{Source Name} &  \multicolumn{2}{c}{$\sigma_{\tau}$} & \colhead{} & \multicolumn{2}{c}{$<$T$_{s}>$ (K)} & \colhead{Notes\tablenotemark{a}} \\
\cline{3-4}\cline{6-7}
\colhead{} & \colhead{} & \colhead{This Work\tablenotemark{b}} & \colhead{D00\tablenotemark{c,d}} & \colhead{} &\colhead{This Work} & \colhead{D00\tablenotemark{d}}
} 
\startdata
0038-7422 & 003824-742212 & 0.02 & 0.041 & & 206.6$\pm$20.3 & 470 & S, N \\
0040-7300 & 003754-725156 & 0.09 & 0.092 & & 128.7$\pm$19.1 & $>$ 240 & S \\
0041-7146 & 003939-714141 & 0.07 & 0.165 & & $>$ 247.4 & $>$ 49 & \\
0041-7146 & 003947-713735 & 0.08 & 0.123 & & $>$ 146.6 & $>$ 47 & \\
0041-7146 & 004047-714559 & 0.01 & 0.019 & & 805.7$\pm$266.1 & $>$ 317 & \\
0048-7412 & 004808-741205 & 0.10 & 0.119 & & 95.9$\pm$13.6 & $>$ 233 & S \\
0053-7313 & 005238-731245 & 0.09 & 0.104 & & 200.8$\pm$12.8 & 211 & \\
0054-7227 & 005218-722708 & 0.06 & 0.063 & & 271.8$\pm$42.0 & $>$ 526 & S \\
0054-7227 & 004956-723554 & 0.13 & 0.140 & & 173.4$\pm$18.2 & 208 & \\
0054-7227 & 005337-723144 & 0.16 & 0.137 & & 185.0$\pm$20.5 & 688 & N \\
0056-7107 & 005611-710707 & 0.02 & 0.012 & & 236.4$\pm$34.8 & 316 & N \\
0056-7107 & 005652-712300 & 0.10 & 0.177 & & 92.8$\pm$10.7 & $>$ 123  & S, N \\
0058-7413 & 005732-741243 & 0.01 & 0.018 & & 239.3$\pm$52.7 & 411 & N \\
0101-7138 & 010029-713826 & 0.05 & 0.069 & & 194.5$\pm$15.8 & 344 & N \\
0110-7135 & 010932-713452 & 0.17 & 0.088 & & 88.5$\pm$14.5 & $>$ 331 & S \\
0110-7227 & 011005-722647 & 0.03 & 0.060 & & 144.0$\pm$16.5 & 390 & S, N \\
0110-7227 & 011035-722807 & 0.12 & 0.189 & & 64.6$\pm$ 6.8 & $>$ 249 & S \\
0111-7314 & 011056-731404 & 0.11 & 0.147 & & 126.3$\pm$11.2 & 322 & S \\
0111-7314 & 011049-731428 & 0.03 & 0.022 & & 252.3$\pm$23.4 & 309 & N \\
0111-7314 & 011432-732143 & 0.23 & 0.087 & & 104.2$\pm$10.6 & 188 & high noise \\
0115-7322 & 011628-731438 & 0.07 & 0.129 & & 197.7$\pm$14.5 & 186 & \\ 
\enddata
\tablenotetext{a}{S = shallow and wide features; N = narrow and deep features}
\tablenotetext{b}{in spectra smoothed to $0.6$ km s$^{-1}$ with $0.2$ km s$^{-1}$ wide channels}
\tablenotetext{c}{in spectra with $0.845$ km s$^{-1}$ wide channels}
\tablenotetext{d}{D00 refers to \citet{dic00}}
\end{deluxetable*}




\bibliography{papers}



\end{document}
